\documentclass[conference]{IEEEtran}
\IEEEoverridecommandlockouts

\usepackage{amsmath,amsfonts,amssymb}
\usepackage{algorithmic}
\usepackage{algorithm}
\usepackage{array}
\usepackage{graphicx}
\usepackage{booktabs}
\usepackage{multirow}
\usepackage{tabularx}
\usepackage{pifont}
\usepackage{makecell}
\usepackage{diagbox}
\usepackage{enumitem}
\usepackage{xcolor}
\usepackage{cite}
\usepackage[hidelinks]{hyperref}
\usepackage{cleveref}

\newcolumntype{C}{>{\centering\arraybackslash}X}
\usepackage{physics}
\usepackage{tikz}
\usepackage[normalem]{ulem}
\usetikzlibrary{positioning, shapes.geometric}
\tikzset{
    red gate/.style={color=red, fill=red!20},
    green gate/.style={color=green!70!black, fill=green!20},
    blue gate/.style={color=blue, fill=blue!20}
}
\usepackage{amsthm}
\theoremstyle{plain}
\newtheorem{theorem}{Theorem}[section]

\theoremstyle{definition}
\newtheorem{definition}[theorem]{Definition}

\theoremstyle{remark}

\newcommand{\qknob}{\textsc{Qknob}}
\newcommand{\queko}{\textsc{Queko}}

\newcommand{\tket}{\text{t}\ensuremath{\mathbf{\ket{\text{ket}}}}}

\begin{document}

\title{QROB: Quantifying Realization Overhead in Quantum Compilation via Reverse Construction}

\author{
    \IEEEauthorblockN{Jintao Li$^{1}$, Kaiqi Li$^{2}$, Rui Wang$^{2}$, Yilun Zhao$^{2}$, Kaixuan Huang$^{1}$, Ying Wang$^{2}$, \\ Jialin Zhang$^{2}$, Zheng-An Wang$^{1, \dagger}$, Xiaoming Sun$^{2, \dagger}$, Heng Fan$^{3, 4, 1, 5, \dagger}$ 
    }
    \IEEEauthorblockA{
    $^{1}$ Beijing Key Laboratory of Fault-Tolerant Quantum Computing, \\
    Beijing Academy of Quantum Information Sciences, Beijing, China \\
    $^{2}$ Institute of Computing Technology, Chinese Academy of Sciences, Beijing, China \\
    $^{3}$ Beijing National Laboratory for Condensed Matter Physics, \\
    Institute of Physics, Chinese Academy of Sciences, Beijing, China \\
    $^{4}$ Beijing Key Laboratory
    of Advanced Quantum Technology, Beijing, China \\
    $^{5}$ Hefei National Laboratory, Hefei, China\\
    }
    \IEEEauthorblockA{
    \{lijt, huangkx, wangza\}@baqis.ac.cn, \{likaiqi, wangrui, wangying2009, zhangjialin, \\
    sunxiaoming\}@ict.ac.cn, zyilun8@gmail.com, hfan@iphy.ac.cn
    }
    \thanks{$^{\dagger}$ Corresponding authors: Zheng-An Wang, Xiaoming Sun, and Heng Fan.}
    \thanks{Accepted to the 59th IEEE/ACM International Symposium on Microarchitecture (MICRO 2026). \copyright~2026 IEEE. Personal use of this material is permitted. Permission from IEEE must be obtained for all other uses, in any current or future media, including reprinting/republishing this material for advertising or promotional purposes, creating new collective works, for resale or redistribution to servers or lists, or reuse of any copyrighted component of this work in other works.}
}
\maketitle

\begin{abstract}
Quantum compilation reconciles a program’s idealized interaction topology with hardware locality constraints, yet evaluations at scale lack calibrated references for realization overhead. We present QROB, a scalable reverse-construction methodology that generates compilation instances backward from directly realizable configurations, retaining the inverse paths as feasible, compiler-independent references. QROB provides a common evaluation substrate for NISQ SWAP routing and fault-tolerant lattice-surgery scheduling, while extending its reference-preserving principle to capacity-constrained quantum memory-access scheduling.

Across systems ranging from 9 to 156 qubits, evaluations highlight QROB’s utility as both a diagnostic benchmark and a data source. First, for compiler characterization, QROB reveals substantial realization gaps in existing tools, with NISQ compilers incurring up to 24.1$\times$ the reference SWAP cost and fault-tolerant compilers requiring up to 7.0$\times$ the reference makespan. Second, as a supervision source for data-driven compilation, a router trained on QROB references outperforms Qiskit SABRE on 84.8\% of real-world application circuits. Finally, on real hardware, QROB reference realizations achieve a median mirror-circuit survival rate $1.65\times$ that of full Qiskit O3 compilations across three 156-qubit IBM Heron-r2 processors, demonstrating that closing algorithmic compilation gaps translates directly into physical fidelity gains.
\end{abstract}


\section{Introduction}
\label{sec:introduction}

Quantum computing is approaching an inflection point. Recent progress in suppressing errors below the fault-tolerant threshold~\cite{google2023suppressing, bluvstein2024logical,bravyi2024high} and demonstrating early logical-qubit operations~\cite{bluvstein2024logical, acharya2024quantum} indicates that the field is beginning to transition from noisy experimental devices to systems capable of fault-tolerant execution. As this transition unfolds, physical resource efficiency becomes the dominant systems bottleneck. The underlying error-correction protocol determines the baseline cost of protected computation, but compilation determines how much additional physical overhead is required to realize a hardware-agnostic program under machine locality constraints. As a result, physical resource efficiency is shaped not only by the fault-tolerant execution model itself, but also by how the compiler maps a hardware-agnostic program onto a constrained machine~\cite{li2019tackling, chong2017programming, litinski2019game}.

This compiler dependence arises from a deeper abstraction gap: quantum algorithms are expressed in a hardware-agnostic circuit representation that does not expose the true machine-level cost of execution~\cite{nielsen2010quantum}. In classical computing, source programs do not explicitly encode the execution cost of cache-line movement, register allocation, or instruction scheduling; those costs emerge only at lower abstraction layers. Quantum computing exhibits an analogous gap. 
The root cause is a mismatch between the \emph{idealized interaction topology} assumed by the program and the \emph{hardware locality constraints} imposed by the target platform. To build intuition for a classical computer architect, imagine a CPU where Arithmetic Logic Units (ALUs) can only perform operations between \emph{physically adjacent} registers. If a program requires an operation between two distant registers, the compiler must explicitly insert a sequence of data-swapping instructions to route them together before the operation can occur. Superconducting-qubit processors, for example, face an analogous constraint: their sparse coupling graphs permit direct two-qubit gates only between hardware-connected qubits~\cite{li2019tackling,siraichi2018qubit}. Fault-tolerant surface-code architectures impose a related locality constraint at the logical-patch level.

These architectural constraints determine which operations are directly realizable and which require costly mediation. On near-term devices, they are captured by the coupling map, so nonlocal interactions incur SWAP-based routing overhead~\cite{li2019tackling,siraichi2018qubit}. In lattice-surgery-based fault-tolerant architectures, patch layouts and routing resources constrain logical interactions, so nonlocal operations incur additional space--time overhead~\cite{litinski2019game,horsman2012surface,molavi2025wisq}. Beyond geometric routing, some fault-tolerant architectures separate a storage-efficient memory region from a smaller compute region, introducing a capacity-constrained logical-qubit residency-management problem~\cite{viszlai2026qsieve,stein2025hetec,kobori2025lsqca}.
These constraints persist across near-term and fault-tolerant systems, although the mechanisms and costs of mediation differ.

Bridging this mismatch is difficult in general, because optimal compilation under locality constraints remains NP-hard~\cite{botea2018complexity, siraichi2018qubit}. This difficulty has motivated a broad spectrum of compilation approaches, including exact formulations for small instances~\cite{tan2020optimal, wille2019mapping, lin2023scalable}, scalable heuristics for practical Noisy Intermediate-Scale Quantum (NISQ) routing~\cite{li2019tackling, cheng2024robust, zou2024lightsabre}, learning-based methods that infer routing policies from data~\cite{tang2024alpharouter,pozzi2022using}, specialized frameworks for fault-tolerant lattice-surgery compilation~\cite{litinski2019game,horsman2012surface,molavi2025wisq}. 
Across these regimes, however, evaluating solution quality at practical scale remains difficult because the achievable optimum is generally unknown.

One common limitation of existing quantum compilation approaches is that they are developed without calibrated reference points for the achievable optimum. 
In systems optimization, progress is most informative when one can reason about the distance to a principled bound rather than relative improvement alone. For example, classical kernel optimization is guided by performance ceilings like the roofline model~\cite{williams2009roofline}. Although quantum analogues like the quantum hardware roofline~\cite{kalloor2024quantum} exist, they are fundamentally \emph{compiler-dependent}. Because their evaluation relies on specific synthesis and compilation workflows to determine algorithmic implementation metrics, their ceilings are fixed per algorithmic workload based on existing compiler outputs. This dependency makes them somewhat circular for judging the true quality of a compiler itself. What remains missing is the \emph{compiler-independent} complement to these models. Such a complement would provide a calibrated reference for the minimum realization overhead required to map an ideal program onto constrained hardware.

In response, we present \textbf{QROB} (Quantum Realization Overhead Benchmark), a reference-preserving reverse-construction framework for quantifying realization overhead under quantum architecture constraints. 

The key insight is a computational asymmetry between forward optimization and reverse construction. Consider a Rubik's Cube: finding a shortest solution to an arbitrary scramble requires expensive search, whereas a designer who starts from the solved state and applies a controlled sequence of moves obtains a valid return path simply by reversing that sequence, although the path is not necessarily shortest.

QROB exploits the same asymmetry in quantum compilation. It starts from a circuit that is directly realizable under the target constraints and applies influence-guided scrambling to construct a more challenging instance. Recording the inverse transformations yields a feasible reference realization and its cost without solving the intractable forward optimization problem. This reference is reproducible and independent of the compiler under evaluation, but it is not assumed to be globally optimal; we assess its quality against exact solvers on tractable instances and provide certification under restricted constructions.

At its core, QROB operates at the level of interaction-locality reconciliation and applies uniformly to NISQ routing and fault-tolerant lattice-surgery scheduling. We additionally test whether its higher-level reference-preserving principle extends beyond geometric routing through a capacity-constrained memory-access case study. 

Our contributions can be summarized as follows:
\begin{itemize}[leftmargin=*, itemsep=2pt]
    \item \textbf{QROB, a unified and scalable reverse-construction methodology.} We present a methodology that retains feasible reference realizations while generating benchmark instances from 9 to 900 qubits in sub-second time (\S\ref{sec:methodology}).

    \item \textbf{A practical tool for characterizing compiler-reference gaps.} QROB provides calibrated references against which a compiler's realization overhead can be rigorously measured, enabling evaluation of performance-scalability tradeoffs (\S\ref{sec:eval}).

    \item \textbf{A high-quality data source for learning-based quantum compilation.} QROB generates calibrated references that can be used as supervision for data-driven compilation, providing scalable training targets for learning-based methods that would otherwise lack validated labels (\S\ref{sec:nn}).
\end{itemize}

Our evaluation demonstrates the value of QROB as both a characterization tool and a data source. Qiskit SABRE incurs $7.2\times$ the QROB reference SWAP count on average, with the ratio growing from $3\times$ on 16-qubit systems to $24\times$ on 156-qubit processors. In the fault-tolerant regime, the evaluated lattice-surgery compilers incur $6.3$--$7.0\times$ the QROB reference execution time. QROB-Mem further provides a retained reference for capacity-constrained memory-access scheduling, enabling quantitative evaluation of scheduling policies in this setting.

As a data source, compilers trained on QROB references outperform SABRE on $84.8\%$ of real circuits, whereas an identical model trained on adversarial benchmark labels achieves $94.7\%$ validation accuracy yet fails on $75\%$ of real workloads. The compiler-reference gap also translates into hardware performance: with full \texttt{Qiskit} O3 enabled on three 156-qubit IBM Heron-r2 processors, QROB reference realizations achieve $1.65\times$ the median mirror-circuit survival of O3's own compilations, with the advantage widening as system size increases.

\section{Background}
\label{sec:background}

Quantum circuits specify the logical qubits that must interact without prescribing how those interactions are realized on physical resources. Each program therefore induces an \emph{interaction topology}, while the target architecture imposes connectivity, spatial, or capacity constraints on its realization~\cite{siraichi2018qubit}. Reconciling these program requirements with architectural constraints is a central challenge in quantum compilation. The core mediation mechanisms considered in this work differ between NISQ SWAP routing and fault-tolerant lattice-surgery scheduling; compute/memory-separated FTQC architectures provide a distinct capacity-constrained extension considered in \S\ref{subsec:memory_access}.

\subsection{NISQ and FTQC Routing}\label{subsec:nisq_vs_ftqc}

\begin{figure}[htbp]
    \centering
    \includegraphics[width=0.96\columnwidth]{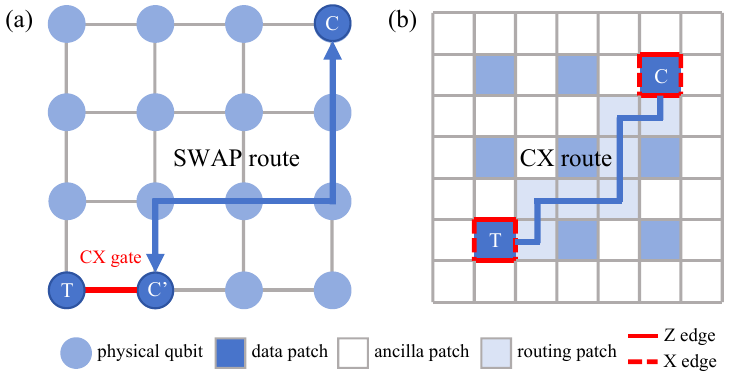}
     \caption{Compilation of a nonlocal $CX$ gate under NISQ and FTQC architecture constraints. $C$ and $T$ denote the Control and Target respectively. (a) On a sparse NISQ coupling graph, a sequence of SWAPs along the blue route moves the control state from its initial location $C$ to $C'$, where a native $CX$ can be executed with the now-adjacent target (red edge). (b) On a schematic Square-Sparse surface-code layout, the data patches remain stationary. Blue squares denote occupied data patches, white squares denote idle ancilla patches, and the light-blue squares denote routing ancilla temporarily allocated to form the lattice-surgery channel. The channel connects a horizontal $Z$ boundary of the control patch to a vertical $X$ boundary of the target patch. Thus, both regimes mediate the same nonlocal logical interaction, but NISQ incurs state movement and additional gates, whereas FTQC incurs temporary spatial occupation and scheduling cost.}
    \label{fig:nisq2ftqc}
\end{figure}

To fully contextualize the compilation bottleneck, it is necessary to distinguish how Noisy Intermediate-Scale Quantum (NISQ) and Fault-Tolerant Quantum Computing (FTQC) paradigms resolve these hardware connectivity constraints, as illustrated in Fig.~\ref{fig:nisq2ftqc}.

\noindent\textbf{NISQ Routing: State Permutation via SWAPs.}
In the NISQ paradigm, each logical variable is directly mapped to a single physical qubit. To execute operations between nonadjacent qubits on the sparse hardware topology, the compiler must actively move quantum states by inserting a sequence of physical \emph{SWAP gates} along a routing path. This specific challenge is formally defined as the \emph{Qubit Routing Problem (QRP)}~\cite{cowtan2019qubit, li2019tackling, siraichi2018qubit, zulehner2018efficient}. The primary compilation overhead is the total number of inserted SWAPs and the resulting circuit depth, which directly exacerbate decoherence.

\noindent\textbf{FTQC Routing: Static Patches and Channel Allocation.}
FTQC fundamentally shifts this abstraction. A logical qubit is encoded across many physical qubits and represented as a macroscopic 2D region known as a \emph{patch}. In the static-patch lattice-surgery model considered here, logical patches are assigned fixed geometric coordinates on the surface-code lattice and remain stationary during execution.

When two non-adjacent logical patches must interact, they perform \emph{lattice surgery}. Instead of physically displacing the data patches, the compiler dynamically allocates a contiguous sequence of vacant intermediate space to establish a temporary interaction medium between the source and target patches. These intermediate spaces are referred to as \emph{routing ancillae}~\cite{litinski2019game, horsman2012surface,fowler2012surface}. The broader challenge of optimally placing these static patches and scheduling such dynamic lattice-surgery channels is formally defined as the \emph{Surface Code Mapping and Routing (SCMR)} problem~\cite{molavi2025wisq}.
Once a multi-cycle lattice surgery operation terminates, the allocated \emph{routing ancillae} are explicitly freed and returned to the pool of available resources.

\subsection{Extension Setting: Quantum Memory-Access Scheduling}
\label{subsec:memory_access}
Several fault-tolerant architecture proposals distinguish a storage-efficient memory region from a smaller, more capable compute region, although their code organizations, supported operations, and physical transfer mechanisms differ~\cite{viszlai2026qsieve,stein2025hetec,kobori2025lsqca}. In the common abstraction considered here, most logical qubits remain idle in memory, whereas $T$ and $CX$ operations are executed in the compute region. A $T$ gate requires its operand to be compute-resident, while a $CX$ gate requires both operands to be compute-resident. Rather than modeling architecture-specific transfer paths, QROB-Mem represents this compute--memory data movement through updates to a capacity-constrained resident set. Throughout, we track a \emph{resident set} $R_t$, the logical qubits currently held in the compute region, whose number is bounded by the compute-region capacity $K$. When a gate requires a nonresident operand, that operand is admitted into the compute region and, since the region operates at capacity, one current resident is simultaneously evicted back to memory; we count this paired admission--eviction as one \emph{compute--memory exchange}, the unit of data-movement cost used throughout this work. Two gates acting on a common logical qubit must retain their relative order, while gates sharing no qubit may be freely reordered; these dependency constraints form the gate-dependency directed acyclic graph (DAG), whose topological orders are exactly the legal execution orders of the circuit. Different execution orders change whether the uses of each qubit cluster together or spread apart in time, and hence the number of exchanges required; the compiler must therefore jointly choose an execution order and a resident-set trajectory that minimize compute--memory exchanges. This joint ordering-and-residency problem closely parallels classical register allocation, whose general formulations are computationally intractable~\cite{sethi1975register}. We refer to this abstract problem as \emph{quantum memory-access scheduling}.

\subsection{Evaluation Metric} \label{subsec:metrics}

To measure the realization overhead across fundamentally different architectures under a single framework, we formally define a generalized implementation cost.

\begin{definition}[Implementation Cost]
\label{def:impl_cost}

Given a circuit $C$, an architecture constraint model $G_C$, a valid realization strategy $\sigma$, and a cost metric $f$ (e.g., SWAP count, execution time, space--time volume, or compute--memory data movements), the implementation cost is
\begin{equation}\mathcal{C}_f(C,G_C;\sigma)=f(\sigma).
\end{equation}
The optimal implementation cost is
\begin{equation}\mathcal{C}_f^*(C,G_C)=\min_{\sigma\in\Sigma(C,G_C)}\mathcal{C}_f(C,G_C;\sigma),
\end{equation}
where $\Sigma(C,G_C)$ denotes the set of valid realizations of $C$ under the constraints specified by $G_C$.\end{definition}

This definition accommodates NISQ routing, fault-tolerant lattice-surgery scheduling, and quantum memory-access scheduling as architecture-constrained realization problems, with $f$ specifying the setting-specific realization cost:

\textbf{Near-Term (NISQ) Regime.}
For SWAP-based routing, $G_C$ specifies the physical coupling map, and $f$ counts the number of inserted SWAP gates. For example, if logical qubits $q_0$ and $q_4$ are initially placed at opposite ends of a five-qubit linear chain, making them adjacent requires at least three SWAPs. Without access to the optimal cost $\mathcal{C}_f^*$, one cannot determine whether a heuristic solution using five SWAPs is near-optimal or substantially suboptimal.

\textbf{Fault-Tolerant (FTQC) Regime.}
For lattice-surgery scheduling, $G_C$ specifies the surface-code layout and the routing resources available for mediating nonadjacent operations. We instantiate $f$ primarily as \emph{execution time}, measured in lattice-surgery scheduling steps, to capture serialization caused by routing conflicts. As a secondary metric, we report \emph{space--time volume (STV)}, where space accounts for both logical patches and routing ancillae, and time records how long these resources remain occupied.

\textbf{Extension: Quantum Memory-Access Regime.}
For the QROB-Mem abstraction, $G_C$ specifies the compute and memory regions, the compute-region capacity $K$, and the initial resident set, and $f$ counts the compute--memory exchanges defined in \S\ref{subsec:memory_access}.

\subsection{Problem Statement}
\label{subsec:problem}

Determining the optimal implementation cost $\mathcal{C}_f^*$ is NP-hard for the core routing problems considered here, and exact methods are generally limited to small instances~\cite{botea2018complexity,lin2023scalable}. Consequently, compiler evaluations at practical scales often lack reference solutions whose quality can be independently assessed.

This work addresses the following problem: Given an architecture constraint model $G_C$ and a cost metric $f$, construct a set of benchmark instances $\{(C_i,R_i,M_i)\}$, where $R_i$ is a retained feasible realization cost and $M_i$ records its construction mode and validation status. The benchmark should (1) provide reference costs calibrated against exact solvers on tractable instances and, where applicable, certified under restricted constructions; (2) generate instances in polynomial time at practical system sizes; and (3) instantiate reference-preserving reverse construction for physical-qubit routing and fault-tolerant lattice-surgery scheduling, while testing its extensibility beyond geometric routing through capacity-constrained memory-access scheduling.

\section{Methodology}
\label{sec:methodology}

\begin{figure*}[htbp]
    \centering
    \includegraphics[width=0.94\textwidth]{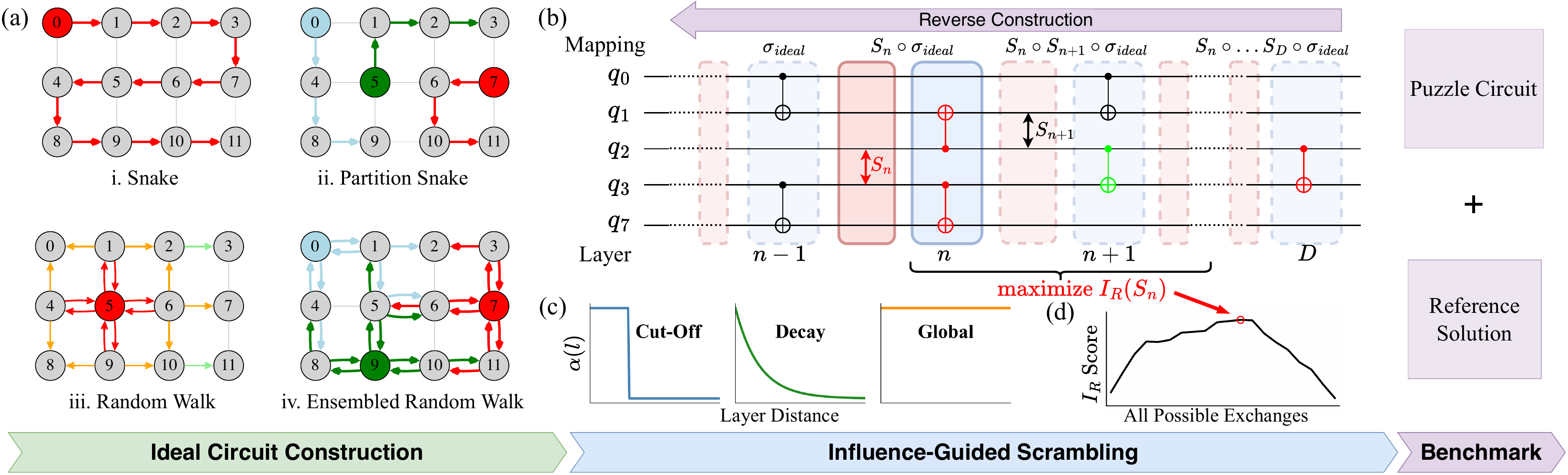}
    \caption{The reverse construction workflow of QROB. (a)~Four strategies for ideal circuit construction. (b)~Influence-guided scrambling at layer $n$: the red arrow represents the candidate exchange operator $S_n$; red and green CX gates denote operations that become nonrealizable and realizable under the new mapping. (c)~Three $I_R$ weighting configurations modeling different planning horizons. (d)~Selection of the highest-scoring exchange operator $S_n^*$ under $I_R$.}
    \label{fig:circuit_construction}
\end{figure*}

QROB is built on a reference-preserving view of benchmark generation. Rather than treating the reference solution as something to be recovered after a circuit is fixed, QROB records the reference solution during circuit generation. This turns reference generation from a post-hoc search problem into a scalable reverse-construction problem. For its two core geometric-routing settings, QROB exploits this asymmetry through a two-stage pipeline (Fig.~\ref{fig:circuit_construction}). First, \textbf{Ideal Circuit Construction} produces a circuit that is directly realizable on the target platform, using one of four strategies ($\mathcal{S}_S,\mathcal{S}_P,\mathcal{S}_R,\mathcal{S}_E$). Second, \textbf{Influence-Guided Scrambling} perturbs the qubit mapping using configurable weighting functions (Cut-Off, Decay, and Global), while retaining the inverse perturbation sequence as a feasible reference. This exchange-and-route construction applies to both physical-qubit routing and fault-tolerant lattice-surgery scheduling. \S\ref{subsec:qrob_mem} then instantiates the higher-level reference-preserving principle beyond geometric routing through a capacity-constrained resident-set construction rather than a geometric exchange trajectory.

\subsection{Preliminaries}
\label{subsec:prelim}

\textbf{Physical and Logical Qubits.}
To systematically characterize the spatial allocation of quantum workloads onto constrained hardware topologies, we first formalize the mapping abstraction between logical and physical qubits.
We model a target platform using a weighted connectivity graph $G_C=(P,E,W)$, where $P=\{p_1,p_2,\dots\}$ is the set of physical qubits, corresponding to individual physical qubits in NISQ devices and surface code patches in FTQC, and $E \subseteq P \times P$ denotes the set of direct connections between nodes. The edge-weight set $W=\{w_{i,j}\}$ captures platform-dependent attributes of the corresponding connections, such as gate fidelity in NISQ or routing cost in FTQC. Let $Q=\{q_1,q_2,\dots\}$ denote the set of logical qubits. A mapping is a bijection $\sigma: Q \rightarrow P$ that assigns each logical qubit to a unique node.
Intuitively, this bijection strictly defines the spatial binding of logical qubits to hardware execution sites across the target topology.

\textbf{Quantum Circuit.}
We mathematically represent the target quantum algorithm as a temporal dependency chain of unitary transformations.
A quantum circuit is defined as an ordered sequence of $n$ quantum gates:
\begin{equation}
\label{eq:ideal_circuit}
C = g^{(n)} \circ \dots \circ g^{(2)} \circ g^{(1)},
\end{equation}
where each $g^{(t)}$ is a quantum gate and $\circ$ denotes successive application. The ordered gate set is $\mathcal{G}(C)=\{g^{(1)}, g^{(2)}, \dots, g^{(n)}\}$.

\textbf{Mapping and Exchange Group.}
To rigorously model the state space of layout reconfigurations necessitated by hardware connectivity constraints, we formulate the routing process as algebraic permutations over a restricted topological graph.
All possible mappings form a mapping space $\Sigma = \{\sigma_0,\sigma_1,\sigma_2,\dots\}$. We consider exchange operations $S(p_i,p_j)$ from an exchange set $\mathcal{S}$, where each $S(p_i,p_j)$ swaps the logical assignments of sites $p_i$ and $p_j$. Under this representation, the reverse-construction process can be viewed as a walk on the restricted Cayley graph $G(\Sigma,\mathcal{S}')$, where exchanges are restricted to connected site pairs on the connectivity graph:
\begin{equation}
\mathcal{S}'=\{S(p_i,p_j)\mid S(p_i,p_j)\in\mathcal{S},\ (p_i,p_j)\in E(G_C)\}.
\end{equation}
From a compilation perspective, traversing this restricted graph mathematically bounds the spatial routing overhead (e.g., SWAP insertions) required to resolve algorithmic-hardware topological discrepancies.

\subsection{Ideal Circuit Construction}
\label{subsec:ideal}

\textbf{Construction Principle.}
The reverse-construction paradigm fundamentally requires a zero-overhead ground truth as its starting point. To achieve this, we first generate an \emph{ideal circuit} $C_{ideal} = \prod_{t=1}^n g^{(t)}$. The macroscopic construction principle is straightforward yet strict: every constituent two-qubit gate $g^{(t)}$ acting on logical qubits $q_i, q_j$ must be directly realizable on the hardware topology without any routing mediation. Mathematically, this enforces the constraint that their mapped physical locations must be adjacent, i.e., $(\sigma(q_i), \sigma(q_j)) \in E(G_C)$. By iteratively sampling edges from the hardware graph and appending the corresponding operations, we macroscopically construct a circuit whose realization cost is inherently optimal (i.e., zero additional routing overhead) by construction.

\textbf{Basic Strategy Library.}
At the macroscopic level, the topological distribution of these sampled interactions ultimately dictates the structural characteristics of the generated benchmark. To provide a diverse structural baseline, QROB incorporates a basic strategy library. For instance, the \emph{Snake} and \emph{Partition Snake} strategies greedily extend paths to mimic deep sequential dependency chains and structured concurrency, respectively. Conversely, the \emph{Random Walk} and \emph{Ensembled Random Walk} strategies naturally concentrate interactions around frequently revisited topological regions, successfully creating localized structural bottlenecks and spatially separated interaction clusters purely through stochastic graph traversal.

\textbf{Framework Extensibility and Custom Strategies.}
Crucially, QROB is designed as a highly extensible generative engine rather than a static, hard-coded benchmark suite. The ideal circuit generator acts as a fully decoupled front-end module. Researchers and compiler developers can seamlessly define and plug in arbitrary custom interaction strategies to synthesize workloads with specific entanglement densities, gate distributions, or architectural stress points, thus effortlessly scaling to satisfy highly specialized evaluation needs.

\textbf{Extended Strategy.}
To illustrate this extensibility and provide a proxy for structural alignment, we introduce an \emph{Extended Variational Strategy}. It emulates the block-wise entanglement layers of variational circuits (e.g., QAOA/VQE) by isolating a bounded-degree subgraph (2--3 neighbors) on the hardware grid. These skeletal edges form parallelizable matching layers that are periodically repeated and locally densified, before entering QROB's standard reverse scrambling pipeline.

As demonstrated in our learning-based evaluation (Fig.~\ref{fig:nn_combined}), augmenting the \emph{Basic Strategy} GNN training set with these \emph{Extended Variational} instances improves performance on real-world variational benchmarks. While not a rigorous mathematical proof of structural equivalence, this cross-domain transferability provides strong circumstantial validation. It confirms that QROB's synthetic instances preserve the essential topological signatures and routing bottlenecks of practical quantum algorithms, serving as effective proxies for compiler optimization.

\subsection{Influence-Guided Scrambling}
\label{subsec:disruption}

Having constructed the ideal circuit $C_{ideal}$ with zero implementation cost, the next step is to transform it into a compilation instance with a controlled realization-overhead gap. This transformation is not a random shuffle; it is performed by the influence-guided scrambling process, which systematically disrupts direct realizability while retaining a reference solution for subsequent evaluation.

We first partition $C_{ideal}$ into $D$ layers, denoted $C_1, C_2, \dots, C_D$. The scrambling process begins from the last layer $C_D$, which is initially associated with the ideal mapping $\sigma_{ideal}$ and is therefore directly realizable. At step $D$, we apply an exchange operator $S_D$ to obtain a new mapping $\sigma_D = S_D \circ \sigma_{ideal}$. This perturbation may break the realizability of some two-qubit operations $g_{i,j} \in \mathcal{G}(C_D)$, since the adjacency condition $(\sigma_D(q_i), \sigma_D(q_j)) \in E(G_C)$ is no longer guaranteed.

QROB then proceeds iteratively backward from layer $C_D$ to $C_1$. For a given layer $C_n$, the newly selected exchange operator $S_n$ updates the effective state used by the reverse-construction process. As illustrated in Fig.~\ref{fig:circuit_construction}(b), the scrambling trajectory is therefore determined by the composition of exchange operators selected along this reverse walk on the restricted Cayley graph.

After completing the reverse traversal, the resulting compilation instance is
\begin{equation}
   C_{instance}(\boldsymbol{\sigma}) = C_D(\sigma_{D}) \circ C_{D-1}(\sigma_{D-1}) \circ \dots \circ C_1(\sigma_{1}),
\end{equation}
where the mapping associated with each layer $k$ is
\begin{equation}
   \sigma_k = S_k \circ S_{k-1} \circ \dots \circ S_1 \circ \sigma_{ideal}.
\end{equation}

The central design problem is therefore to select a sequence of exchange operators $\{S_1, \dots, S_D\}$ that maximally increases the realization overhead required to restore direct realizability. To guide this selection, we define a score function $\Delta(S_k, C_l)$ that measures the degradation induced on layer $C_l$ by a candidate exchange $S_k$. These per-layer effects are aggregated into an influence score
\begin{equation}
    I_R(S_k) = \sum_{l=1}^{D} \alpha_{k,l}\,\Delta(S_k, C_l),
\end{equation}
where $\alpha_{k,l}$ specifies how strongly the effect on layer $C_l$ is coupled to the current scrambling decision at step $k$. Under this unified framework, the same reverse-construction process can be instantiated for different routing settings.

From a physical routing perspective, $I_R$ models the cascading routing disruption caused by a physical swap operation. Intuitively, it asks: when a candidate exchange $S_k$ displaces logical qubits on the hardware topology, how much does this movement destroy the required physical adjacency for both current and future operations? The term $\Delta(S_k,C_l)$ measures this localized loss of adjacency within a specific layer $C_l$. Crucially, $\alpha_{k,l}$ defines the compilation lookahead window. Cut-Off models a short-sighted disruption (finite lookahead), Decay applies a gradually diminishing penalty to future layers, and Global treats the entire execution sequence as deeply coupled. Thus, $I_R$ serves as a synthetic scoring function to deliberately inject controlled, cross-layer routing bottlenecks during benchmark generation.

\textbf{NISQ SWAP-based routing.}
On coupling graphs, we instantiate the score in terms of \emph{executability}, defined as the weighted count of directly executable two-qubit gates:
\begin{equation}
    \mathrm{Exec}(C(\sigma)) =
    \sum_{\substack{g_{i,j} \in \mathcal{G}(C)\\(\sigma(q_i),\sigma(q_j)) \in E(G_C)}}
    W(\sigma(q_i),\sigma(q_j)).
\end{equation}
Accordingly,
\begin{equation}
    \Delta(S_k, C_l)=\Delta \mathrm{Exec}(S_k, C_l),
\end{equation}
where
\begin{equation}
    \Delta \mathrm{Exec}(S_k, C_l)
    =
    \mathrm{Exec}(C_l(\sigma)) - \mathrm{Exec}(C_l(S_k \circ \sigma)).
\end{equation}
Larger values correspond to exchanges that more strongly disrupt direct realizability and therefore require greater SWAP-based routing overhead to restore.

\textbf{Fault-tolerant lattice-surgery routing.}
On surface-code lattices, we instantiate the score in terms of conflict pressure over shortest routing paths. For a candidate exchange $S_k$ applied at layer $C_l$, let $\mathcal{P}_{\text{shortest}}$ denote the set of shortest routing paths for the affected logical operation under the updated mapping, and let $\mathrm{Used}$ denote the set of routing ancillae already reserved by other operations in the retained reference assignment. Accordingly,
\begin{equation}
    \Delta(S_k, C_l)=\Delta \mathrm{Conf}(S_k, C_l),
\end{equation}
where
\begin{equation}
    \Delta \mathrm{Conf}(S_k, C_l)=
    \frac{|\{P \in \mathcal{P}_{\text{shortest}} : P \cap \mathrm{Used} \neq \emptyset\}|}{|\mathcal{P}_{\text{shortest}}|}.
\end{equation}
Here, $\Delta \mathrm{Conf}(S_k, C_l)$ measures the conflict pressure induced by $S_k$ on the retained shortest-path structure of layer $C_l$. Larger values correspond to exchanges that make conflict-free shortest-path restoration less accessible and therefore increase the routing contention encountered during lattice-surgery compilation.

At each step, QROB evaluates all admissible candidate exchanges in $\mathcal{S}'$ and selects the one maximizing $I_R$. The reverse application order of the resulting exchange sequence defines the retained reference solution for the generated instance. Algorithm~\ref{alg:qrob_generation} summarizes the full reverse-construction procedure.

\begin{algorithm}[htbp]
\caption{QROB: Reverse Construction with Retained References}
\label{alg:qrob_generation}
\begin{algorithmic}[1]
\REQUIRE Connectivity graph $G_C(P,E,F)$, strategy $\mathcal{S}$, $I_R$ weighting $\alpha$, depth $D$, density $\rho$
\ENSURE Compilation instance $C_{instance}$, reference solution $S_{ref}$
\STATE \textbf{Phase 1: Ideal Configuration Construction}
\STATE $C_{ideal} \leftarrow \textsc{BuildCircuit}(G_C, \mathcal{S}, D, \rho)$
\STATE \textbf{Phase 2: Influence-Guided Scrambling}
\STATE $\sigma \leftarrow \text{IdentityMapping}(P)$; $S_{apply} \leftarrow [\,]$
\FOR{$\ell = D{-}1$ \textbf{down to} $0$}
    \STATE $\text{max\_ir}, \text{best\_exchange} \leftarrow \text{null}$
    \FOR{each exchange $S \in \{(p_i,p_j) \mid (p_i,p_j) \in E\}$}
        \STATE $\text{current\_ir} \leftarrow I_R(S, \alpha)$
        \IF{$\text{current\_ir} > \text{max\_ir}$}
            \STATE $\text{max\_ir} \leftarrow \text{current\_ir}$; $\text{best\_exchange} \leftarrow S$
        \ENDIF
    \ENDFOR
    \IF{$\text{max\_ir} > \epsilon$}
        \STATE $\sigma \leftarrow \text{best\_exchange}(\sigma)$
        \STATE $S_{apply}.\textsc{append}(\text{best\_exchange})$
    \ENDIF
\ENDFOR
\STATE $C_{instance} \leftarrow \textsc{ApplyMapping}(C_{ideal}, \sigma)$
\STATE $S_{ref} \leftarrow S_{apply}.\textsc{reverse()}$
\RETURN $(C_{instance}, S_{ref})$
\end{algorithmic}
\end{algorithm}

\textbf{Controlling Structural Regimes.}
The scrambling process exposes two largely separable aspects of the generated instances. The first is the \emph{reference cost}, determined by the amount of mediation required by the retained reference solution. The second is the \emph{structural regime} induced by the score function, which governs whether the resulting realization-overhead gap remains localized or becomes more globally coupled within the generated instance. The weighting function $\alpha_{k,l}$ controls this latter aspect by shaping how broadly the induced perturbation is coupled across the instance. In this way, QROB can generate instances with comparable reference cost but substantially different structural organization, enabling finer-grained characterization of compilation behavior within the same reverse-construction framework.

\noindent\textbf{Time Complexity:} The overall complexity is $O(D \cdot |E| \cdot d)$, polynomial in circuit and graph size.

\subsection{Extension to Memory-Access Scheduling}\label{subsec:qrob_mem}
QROB-Mem instantiates the reference-preserving principle for the memory-access setting of \S\ref{subsec:memory_access}, operating on the resident set defined there rather than on a geometric mapping. The construction first plants a trajectory of resident sets $R_1,\dots,R_P$, each containing exactly $K$ logical qubits drawn from an active pool of $A>K$ recurring qubits; consecutive sets differ in a small number of members, so qubits leave the compute region and later return. For each stage $i$, it then emits a batch of $CX$/$T$ gates whose operands all lie in $R_i$ and which collectively touch every member of $R_i$. Because every gate in stage $i$ acts only on members of $R_i$, the recorded resident-set trajectory provides a legal full-capacity realization of that stage. Replaying the planted trajectory therefore executes the circuit while admitting only the members entering at each transition, and its total cost of $\sum_{i>1}|R_i\setminus R_{i-1}|$ exchanges is recorded and machine-checked. For example, with $K{=}2$, $R_1=\{q_1,q_2\}$, and $R_2=\{q_1,q_3\}$, the stage-2 gates touch $q_3$, so $q_3$ is admitted and $q_2$ is evicted back to memory---one exchange. During evaluation, qubit labels are permuted, and the scheduler receives only the gate-dependency DAG, the capacity $K$, and the initial resident set $R_0=R_1$; because the DAG admits many legal execution orders, the scheduler must jointly rediscover an execution order and a residency schedule, while the retained trajectory provides a feasible reference of known exchange cost. Unlike core QROB, this extension does not construct spatial routes; it tests whether the same reference-preserving principle applies to a capacity-constrained scheduling problem.

\subsection{Reference Quality}
\label{subsec:optimality}

A central question for QROB is what kind of reference quality it provides and how the resulting gaps should be interpreted. QROB does \emph{not} claim to solve the optimal compilation problem for arbitrary circuits, which remains NP-hard. Instead, it generates references under two complementary modes, each serving a different role in the evaluation pipeline.

\textbf{IR mode.} The influence-guided mode of \S\ref{subsec:disruption}, named for the $I_R$ score it maximizes, is designed to generate structurally diverse and controllably difficult instances for stress-testing and learning. 
Its retained references serve as high-quality heuristic baselines. While they do not theoretically guarantee global optimality for arbitrary scales, on tractable small-scale instances, they are empirically shown to closely track the exact choices of the OLSQ2 solver~\cite{lin2023scalable}.

\textbf{Controlled Distance Disruption (CDD) mode.} 
This mode is intentionally designed as a theoretical validation baseline rather than a representative application workload. By applying exchanges along strictly controlled, collateral-free chains, CDD deliberately sacrifices structural diversity to guarantee mathematically certified optimality at scale, where exact solvers fail. This provides an absolute baseline anchor to verify the correctness of the reverse-construction scaling behavior.

Accordingly, QROB establishes reference quality in two complementary ways: solver-based calibration for IR mode on tractable instances, and construction-based certification for CDD mode at large scale. Comparisons against QROB references should therefore be interpreted conservatively: the references are certified under the restricted structural constraints of CDD mode, and act as nontrivial, optimized baseline paths in IR mode.

\begin{figure}[t!]
    \centering
    \includegraphics[width=\columnwidth]{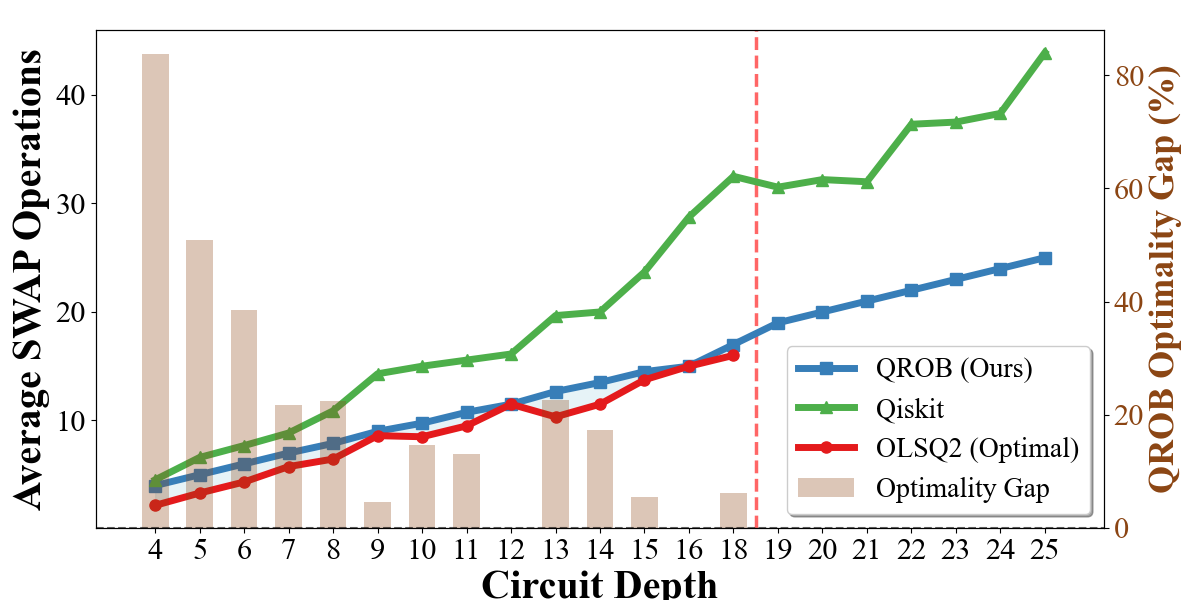}
    \caption{Reference solution quality on a $3\times 3$ grid. QROB (IR mode, blue) closely tracks the OLSQ2 optimal solver (red) across circuit complexities, while \texttt{Qiskit} (green) degrades. Brown bars show QROB's gap relative to OLSQ2 (right axis).}
    \label{fig:optimality_validation}
\end{figure}

\textbf{Solver-based validation.}
On small instances where exact solving is feasible, we calibrate IR-mode references against OLSQ2. As shown in Fig.~\ref{fig:optimality_validation}, IR-mode references closely track the exact optimum across increasing circuit complexity on a $3\times 3$ grid, while \texttt{Qiskit}'s heuristic degrades substantially. This validates QROB's scalable reference-solution generator as closely tracking exact optima on tractable, structurally diverse instances.

\textbf{Construction-based certification via CDD.}
To provide certified references beyond the reach of exact solvers, we introduce a Controlled Distance Disruption (CDD) mode. Unlike the general influence-guided scrambling, CDD applies exchanges along strictly controlled distance-increasing chains under three constraints: monotone distance growth, qubit-disjoint chains, and collateral-free scrambling. Together, these conditions ensure that each disrupted interaction requires an independently necessary reversal cost, so the retained reference is additive and certifiable by construction.

\textbf{Graph-theoretic lower bound.}
For any generated instance, each nonadjacent interaction requires at least $\mathrm{dist}(q_i,q_j)-1$ exchanges. We therefore construct a conflict graph over disrupted interactions, connect interactions that share a qubit, and compute a greedy maximum-weight independent set. This yields a valid lower bound
\[
\mathrm{LB}=\sum_{g\in \mathrm{MWIS}} (\mathrm{dist}(g)-1),
\]
because the selected interactions are qubit-disjoint and their exchange requirements cannot be shared.

\subsection{Scope and Limitations}\label{subsec:scope}

QROB isolates the realization overhead induced by architecture constraints---SWAP-based routing in the NISQ regime, lattice-surgery channel allocation in the FTQC regime, and compute--memory data movement in the QROB-Mem extension---from the rest of the compilation and execution stack. Accordingly, QROB is a benchmark for this isolated layer rather than a full-system model, and its limitations fall into two parts.

\textbf{First, QROB does not model the full compilation and execution stack in any of these settings.} On the NISQ side, we assume input circuits are already optimized by upstream logical synthesis; advanced synthesis techniques, such as phase folding or gate cancellations, are orthogonal to the spatial routing bottleneck and fall outside our scope. On the FTQC side, QROB does not model code-distance-dependent logical errors, syndrome-extraction or decoder latency, factory placement and footprint, distillation throughput, or queuing among concurrent magic-state requests. Full-system analyses can combine QROB's routing estimates with explicit synthesis, error-correction, and distillation models, such as~\cite{sharma2026spacetime}.

\textbf{Second, our magic-state routing evaluation adopts a single simplified supply abstraction.} In \S\ref{sec:ftqcgap}, we model magic-state retrieval from boundary supply locations as traffic on the shared routing fabric, following the abundant-supply abstraction adopted by the evaluated compilers~\cite{molavi2025wisq, beverland2022edpc}. This choice provides a common abstraction under which all evaluated compilers are compared on equal terms, but we do not investigate which factory organization is preferable; as magic-state architectures are still evolving rapidly, this simplified setting may not reflect future designs. The reference makespan is machine-verified under this abstraction; the resulting headroom should therefore be interpreted as a routing-level measurement conditioned on this setting, rather than as a prediction of end-to-end $T$-state production throughput under alternative factory organizations. The reference-preserving construction itself, however, is not tied to any particular supply abstraction and can in principle be re-instantiated as these architectures mature.

\section{Evaluation}
\label{sec:eval}

We evaluate QROB references against seven compilers spanning the compilation stack and across 18 hardware configurations. We first characterize realization overhead on coupling maps and diagnose its structure (\S\ref{sec:gap}), and then extend this core routing analysis to fault-tolerant lattice surgery (\S\ref{sec:ftqcgap}). We next use QROB-Mem as an extension case study to test reference-preserving reverse construction beyond geometric routing (\S\ref{sec:qmas}). Finally, we study QROB references as supervision for data-driven compilation (\S\ref{sec:nn}), validate their structural and hardware relevance (\S\ref{sec:validation}), and directly measure the fidelity gap under full industrial optimization (\S\ref{sec:o3gap}).

\subsection{Experimental Setup}
\label{sec:setup}

The coupling-map compiler comparison (Table~\ref{tab:compiler_gap}) and OLSQ2 calibration experiments were run on an Intel Core i7-14700KF workstation with 64\,GB RAM; learning and fault-tolerant experiments were run on a dual AMD EPYC 9654 server with 1\,TB RAM and NVIDIA RTX PRO 6000 GPUs. For coupling-map compilation, we evaluate five baselines spanning practical routing stacks: \texttt{Qiskit} SABRE (v1.2.4)~\cite{li2019tackling}, \texttt{Cirq} (v1.5.0), \texttt{Tket} (v2.7.0)~\cite{sivarajah2020tket}, \texttt{QMAP} (v2.6.0)~\cite{zulehner2018efficient}, and a greedy baseline. In addition to this standalone-router comparison, \S\ref{sec:o3gap} evaluates the complete \texttt{Qiskit} O3 production pipeline. For fault-tolerant compilation, we evaluate two lattice-surgery compilers: WISQ/DASCOT~\cite{molavi2025wisq}, the dependency-aware compiler of Molavi et al., and EDPC~\cite{beverland2022edpc}, the edge-disjoint-paths compiler.

QROB benchmark suites span 18 hardware configurations: 12 coupling topologies (9--156 qubits) and 6 surface-code lattices (16--256 logical qubits), with four complexity levels per configuration. Fault-tolerant instances are generated by the reverse construction of \S\ref{sec:methodology}, with 8 instances per point. For learning-transfer evaluation, we route held-out real application circuits from QASMBench~\cite{li2023qasmbench} and MQT Bench~\cite{quetschlich2023mqt}; these circuits are used only for final evaluation and never as routing-label sources. A separate structural comparison uses 216 circuits from 10 standard families in Qiskit's circuit library (\S\ref{sec:validation}). Coupling-map compilers are measured primarily in SWAP count relative to the QROB reference; we also report two-qubit depth and compilation time where relevant. Fault-tolerant compilers are measured primarily in execution time, defined as the number of lattice-surgery scheduling steps relative to a dependency-depth reference, with space--time volume reported as a secondary metric.

The main fault-tolerant comparison uses WISQ/DASCOT's square-sparse layout~\cite{molavi2025wisq} and EDPC's native ancilla grid~\cite{beverland2022edpc}, with DASCOT's compact layout as a routing-space control (\S\ref{sec:ftqcgap}).

For ML compilation experiments, we use an EnhancedGATPolicy architecture (GATv2~\cite{brody2021attentive}, 4 layers, 4 attention heads, 47K parameters) with 12-dimensional node features and 6-dimensional edge features. The identical architecture is evaluated under three supervision settings: SABRE heuristic labels, generic QROB retained-reference labels, and family-conditioned QROB labels. The family-conditioned checkpoint uses QROB-generated Extended Variational Strategy instances; held-out QASMBench and MQT Bench circuits are used only for final evaluation and never as routing-label sources. This isolates the effects of label quality and workload conditioning from model capacity. Hardware validation is performed on the Quafu Baihua 156-qubit superconducting processor~\cite{Quafu}, using a 12-qubit high-fidelity sub-topology. The hardware fidelity-gap experiments (\S\ref{sec:o3gap}) additionally use three 156-qubit IBM Heron-r2 processors (\textit{ibm\_kingston}, \textit{ibm\_marrakesh}, \textit{ibm\_fez}) with \texttt{Qiskit} 1.2.4 at \texttt{optimization\_level=3}.

\textbf{Benchmark protocol.}
We release QROB as a frozen evaluation suite, a configurable generator, and a learning suite. Each frozen-suite instance includes its circuit, target coupling map or patch-array layout, generator configuration, random seed, retained reference solution, reference cost, and reference-mode metadata. Compiler versions, settings, and per-instance time budgets are released alongside. Results must report the suite name, target metric, per-instance and aggregate costs, dispersion, completion rate, and all timeouts. Generator-suite results must disclose all construction parameters and be reported separately from frozen-suite compiler rankings. For learning-based use, models train only on QROB-generated instances, while held-out real-circuit families (QASMBench~\cite{li2023qasmbench}, MQT Bench~\cite{quetschlich2023mqt}) and held-out generator configurations are reserved for evaluation.

\subsection{Compilation Gap Characterization}
\label{sec:gap}

\begin{table}[htbp]
\centering
\caption{Coupling-map realization overhead relative to QROB IR-mode references (9--156 qubits; lower is better).}
\label{tab:compiler_gap}
\small
\begin{tabular}{@{}lcccc@{}}
\toprule
\textbf{Compiler} & \textbf{SWAP$\times$} & \textbf{Depth$\times$} & \textbf{Time (s)} & \textbf{Scale$\times$} \\
\midrule
\texttt{Qiskit} SABRE  & \textbf{7.2}  & 15.7 & \textbf{0.10} & \textbf{6.9} \\
\texttt{Tket}           & 12.9 & 52.4 & 5.60 & 18.6 \\
\texttt{Cirq}           & 16.0 & 35.3 & 6.31 & 18.2 \\
\texttt{QMAP}           & 23.0 & 65.7 & 18.5 & 28.0 \\
\midrule
QROB reference          & 1.0  & 1.0  & $<$1 & 1.3 \\
\bottomrule
\end{tabular}
\end{table}

Compilation under locality constraints incurs realization overhead whose magnitude has never been measured against calibrated references. Because QROB references are defined at the level of interaction-locality reconciliation, we can for the first time measure this gap uniformly across 18 hardware configurations---coupling maps from 9 to 156 qubits---evaluating five industrial compilers. 

\begin{figure*}[htbp]
   \centering
   \includegraphics[width=0.88\textwidth]{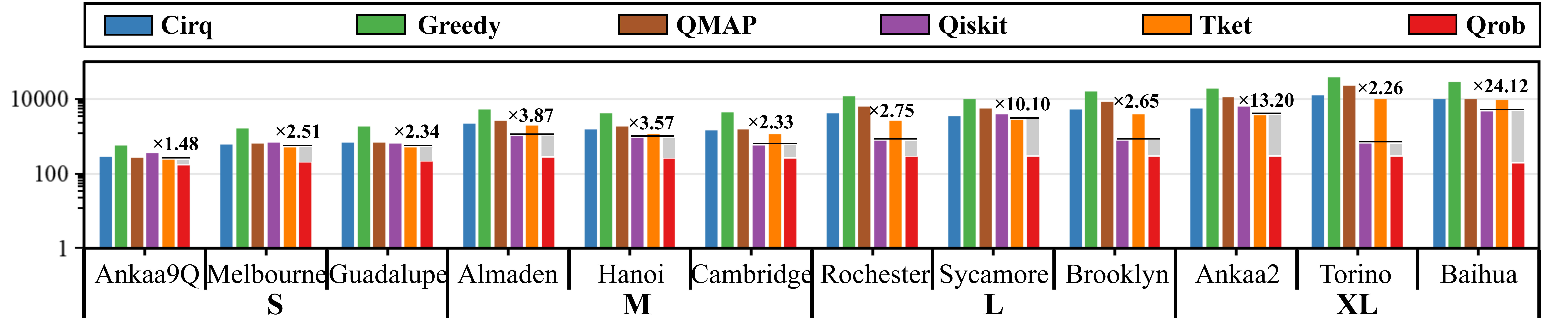}\\
    \begin{minipage}{1.0\linewidth}
		\centerline{\includegraphics[width=0.88\textwidth]{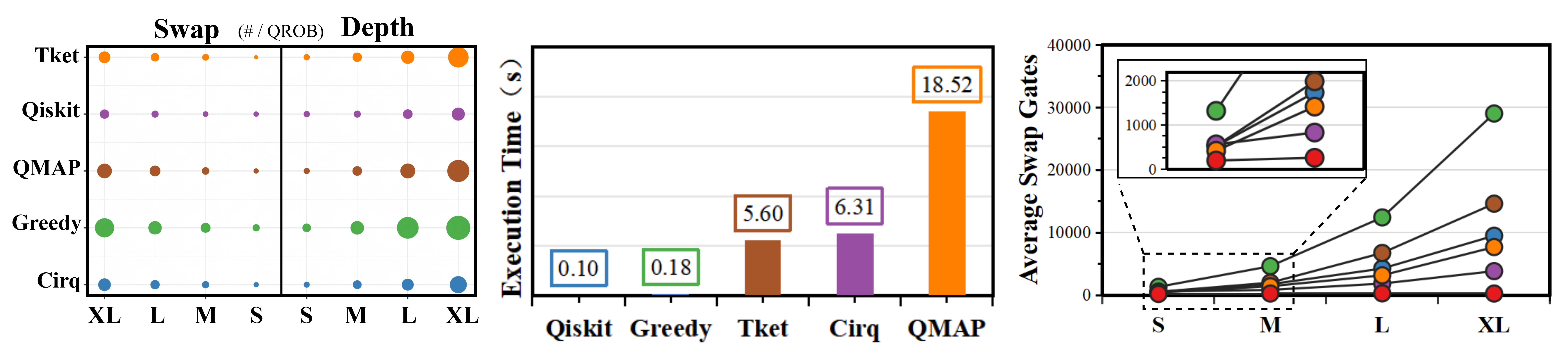}}
	\end{minipage}
   \caption{Realization overhead across 12 coupling-map configurations (9--156 qubits) and four complexity levels. Top: cost ratios relative to QROB references (IR mode). Bottom: execution time and scaling behavior. The cost ratio escalates from $3\times$ at 16 qubits to $24\times$ at 156 qubits.}
   \label{fig:overall}
\end{figure*}

Table~\ref{tab:compiler_gap} and Fig.~\ref{fig:overall} quantify the realization overhead on coupling-map topologies. The gap between current practice and the QROB references is substantial: even the best-performing compiler (\texttt{Qiskit} SABRE) incurs \textbf{7.2$\times$} the reference cost, while \texttt{Tket}, \texttt{Cirq}, and \texttt{QMAP} incur 12.9--23.0$\times$. These gaps were entirely invisible before calibrated references---prior evaluations could only perform \emph{uncalibrated} relative comparisons, with no way to distinguish unavoidable cost from implementation inefficiency. That even \texttt{Qiskit} SABRE---the industry standard with years of engineering investment---is no exception confirms the gap is not an artifact of immature implementations but a fundamental property of the forward-solving paradigm.

\textbf{The overhead ratio grows with scale.} A consistent scaling pattern emerges: as system size grows, the gap widens rather than narrows. On coupling-map topologies, \texttt{Qiskit}'s overhead grows from $3\times$ at 16 qubits to $24.1\times$ at 156 qubits---an $8\times$ increase in the ratio over a $10\times$ increase in system size. This headroom also persists when \texttt{Qiskit}'s full O3 optimization is enabled for both configurations: the ratio still reaches ${\sim}10\times$ at 119 qubits and converts into a measured hardware fidelity loss (\S\ref{sec:o3gap}).

\textbf{The gap is cross-layer.} The scaling pattern is not specific to one compilation setting. Whether the connectivity graph is a coupling map (NISQ) or a surface-code lattice (FTQC), the overhead ratio grows with system size and resists heuristic optimization. This cross-layer consistency suggests that the underlying cause---the exponential growth of the mapping search space relative to the compiler's planning horizon---is a structural property of locality-constrained compilation rather than an artifact of any particular compiler or hardware platform.

\subsection{Lattice-Surgery Compilation}\label{sec:ftqcgap}

\begin{figure}[htbp]
    \centering
    \includegraphics[width=\columnwidth]{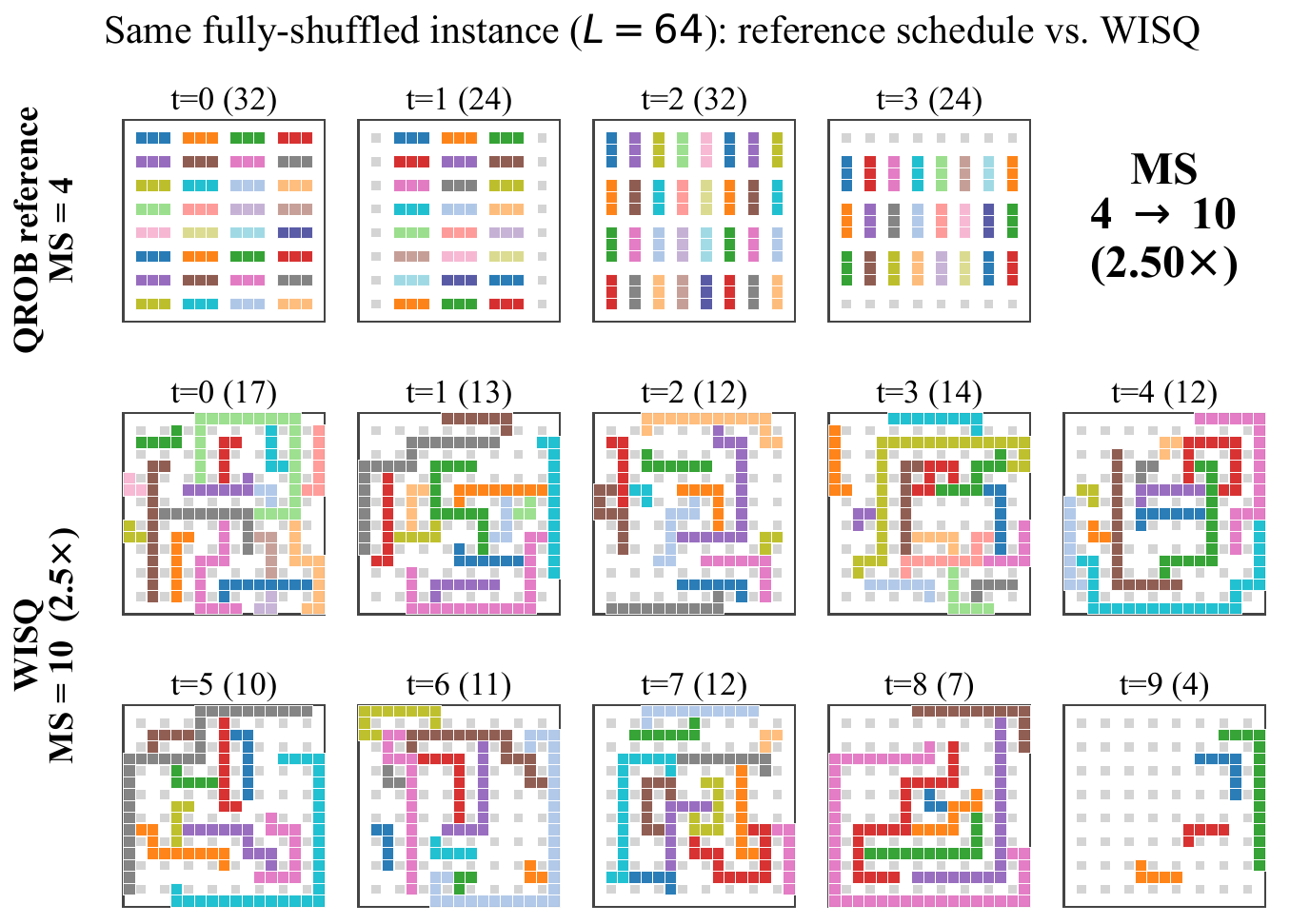}
    \caption{QROB (IR mode) reference schedule vs.\ WISQ on the same fully-shuffled $L{=}64$ instance. Each panel shows the merges executing in one parallel step (count in parentheses); the placement is fixed across steps. The reference placement completes all interactions in $4$ steps, whereas the WISQ placement requires $10$ steps, corresponding to a $2.5\times$ makespan ratio determined by placement quality.}
    \label{fig:ftqc_schedule}
\end{figure}

In the fault-tolerant regime the relevant cost is the lattice-surgery makespan, which is governed primarily by patch placement: a placement that keeps interacting patches adjacent allows many merges to execute in parallel within a single step, whereas a poor placement forces them to serialize through shared routing channels. Fig.~\ref{fig:ftqc_schedule} illustrates this on a single $L{=}64$ instance: the QROB reference completes all interactions in 4 parallel steps (24--32 merges per step), while WISQ requires 10 (4--17 per step). With patches fixed, the routing freedom that remains is limited, so this 2.5$\times$ makespan ratio is determined by placement quality.

\begin{figure}[htbp]
      \centering
      \includegraphics[width=\columnwidth]{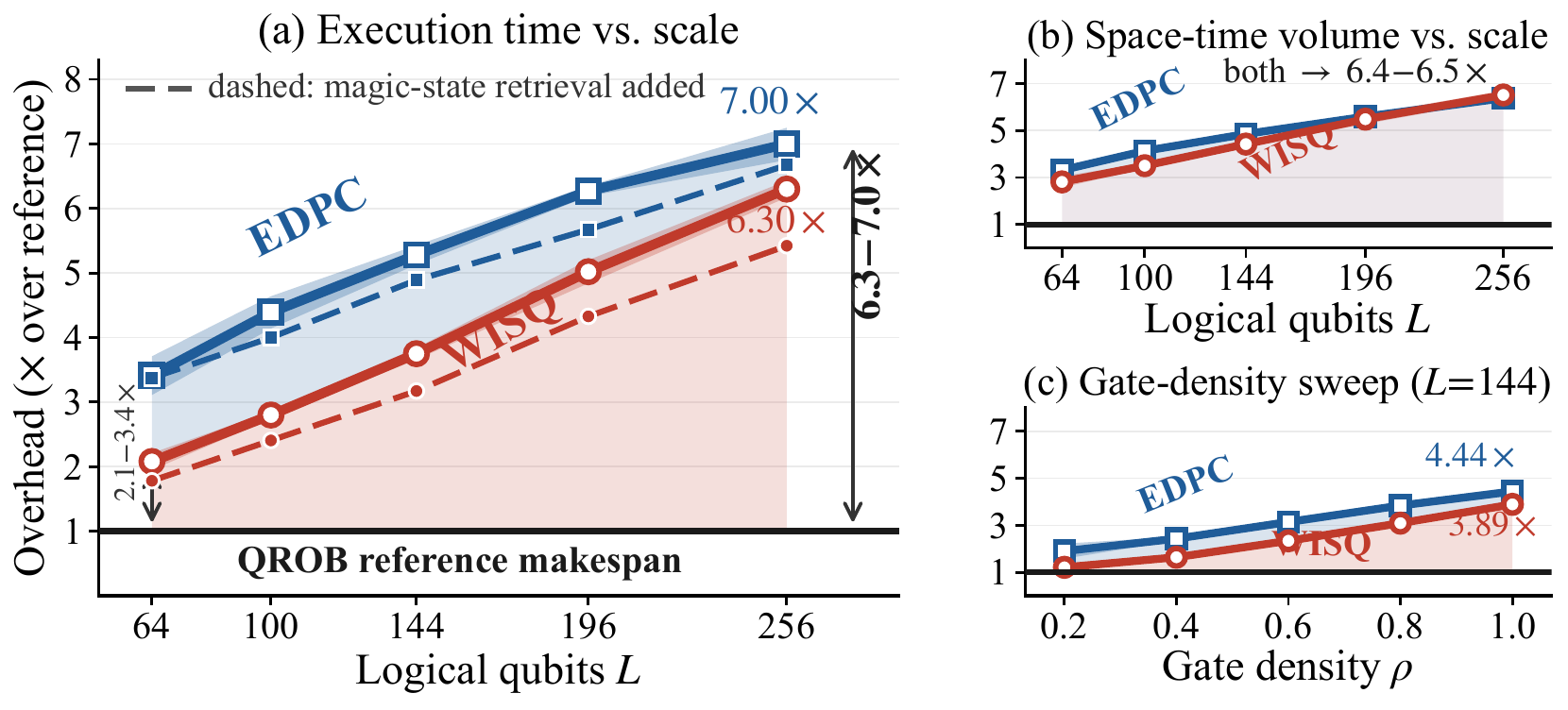}
      \caption{Realization overhead of WISQ and EDPC on lattice-surgery instances, measured against QROB references (IR mode; 8 instances per point; shaded bands show $\pm$1 s.d.). (a)~Execution-time overhead relative to the dependency-depth reference makespan. (b)~Space--time volume overhead, with each compiler measured against its own reference; the two volume models are not directly comparable. (c)~Time overhead as the gate-density knob $\rho$ sweeps at $L{=}144$. WISQ/DASCOT uses the square-sparse layout, and EDPC uses its native ancilla grid. Dashed curves in (a) come from an additional magic-state run in which magic-state retrieval is charged to the same fabric at full $T$ density (\S\ref{sec:ftqcgap}).}
      \label{fig:ftqc_hero}
\end{figure}

Across surface-code lattices from 64 to 256 logical qubits (8 instances per point, generated by the reverse construction of \S\ref{sec:methodology}), the makespan ratio grows with scale: WISQ rises from 2.1$\times$ at $L{=}64$ to 6.3$\times$ at $L{=}256$, and EDPC from 3.4$\times$ to 7.0$\times$ (Fig.~\ref{fig:ftqc_hero}(a)). The same trend holds in space--time volume (Fig.~\ref{fig:ftqc_hero}(b)) and under a tunable gate-density knob (Fig.~\ref{fig:ftqc_hero}(c)).

On DASCOT's compact layout, which halves the routing space, the ratio is larger still---9.2$\times$ at $L{=}64$ to 43.8$\times$ at $L{=}256$. One might expect a compact layout, with its minimal routing space, to leave little headroom; instead the ratio is largest there, because with little room to compensate for a poor placement the makespan becomes even more sensitive to placement quality---the reference still runs most merges in parallel, while a weaker placement is forced onto the single shared corridor.

The instances above are Clifford, so we also charge magic-state retrieval to the same ancilla fabric and test whether the measured headroom persists when $T$ gates are included. In each layer, a controlled fraction $\tau$ of the boundary-ring qubits issues a $T$ request \emph{instead of} joining the CX matching, so the layer count is unchanged and $\tau{=}0$ exactly recovers the construction above. Each $T$ target is boundary-adjacent in the reference placement and its shortest retrieval path is tile-disjoint from that layer's merges, so the reference makespan remains the layer count and is machine-verified. Using their native $T$-gate support, we evaluate both compilers over $L{=}64$--$256$ and four $T$ densities ($160$ instances per compiler; $100\%$ completion). 
At full $T$ density and $L{=}256$, WISQ and EDPC require $5.4\times$ and $6.7\times$ the reference makespan, respectively, with the scaling trend preserved (dashed curves in Fig.~\ref{fig:ftqc_hero}(a)).

Within this supplementary run, $T$ traffic reduces the ratio by only $12$--$18\%$ for WISQ and $4$--$9\%$ for EDPC, with no systematic trend across system scales. The effect remains modest even at $L{=}64$, where $T$-state requests account for $65\%$ of all operations but reduce the ratio by only $12.3\%$. The $\tau{=}0$ results agree with the main Clifford-only baseline within one standard deviation.

\subsection{Extension Case Study: Memory-Access Scheduling}\label{sec:qmas}

We evaluate the set-level QROB-Mem abstraction introduced in \S\ref{subsec:qrob_mem}. A scheduler receives the gate-dependency DAG, compute-region capacity $K$, and initial resident set, and must jointly determine a legal execution order and resident-set trajectory. We evaluate $K\in\{4,8,12,16,20,24\}$ over a system containing 200 logical qubits, with an active pool of $A=2K$ recurring qubits and 20 generation seeds per configuration. The metric counts the compute--memory exchanges defined in \S\ref{subsec:memory_access} and abstracts away internal memory organization, access latency, spatial transfer paths, and magic-state delivery.

\begin{figure}[htbp]
  \centering
  \includegraphics[width=\columnwidth]{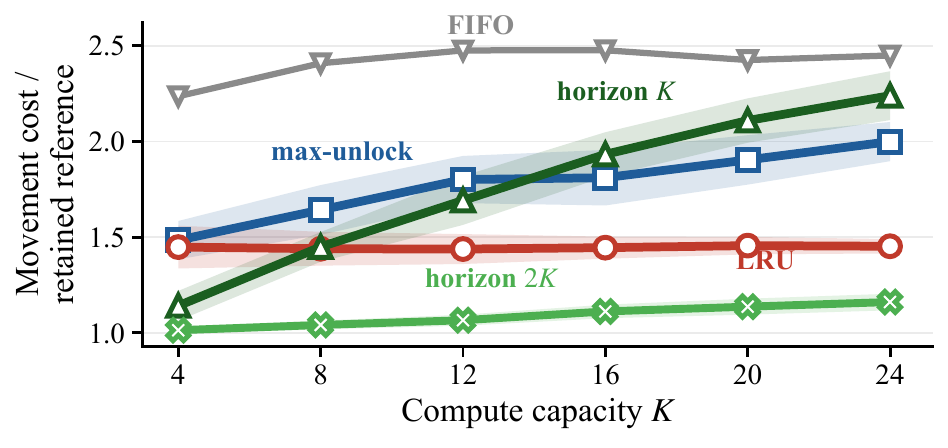}
  \caption{QROB-Mem extension results: compute--memory data-movement cost relative to the retained reference as compute-region capacity $K$ increases ($A=2K$; 20 seeds per point; bands show $\pm1$~s.d.). FIFO and LRU are history-only policies, next-use uses a bounded $K$- or $2K$-gate horizon, and max-unlock operates on the dependency-ready frontier.}
  \label{fig:qmas}
\end{figure}

Fig.~\ref{fig:qmas} compares five policies. FIFO and LRU use admission history, the two next-use policies evict the resident whose next use is farthest within a $K$- or $2K$-gate horizon, and max-unlock admits the qubit that enables the largest number of dependency-ready gates. Lookahead is measured along the original program order, which is a legal topological order of the DAG. The next-use and max-unlock policies are informed by qSIEVE and HetEC, respectively~\cite{viszlai2026qsieve,stein2025hetec}; we did not identify public implementations suitable for direct comparison.

A fixed absolute lookahead does not transfer across compute capacities. The $K$-gate next-use policy increases from $1.14\times$ the retained-reference cost at $K=4$ to $2.24\times$ at $K=24$, whereas the $2K$-gate policy remains within $1.01$--$1.16\times$. In particular, an eight-gate horizon is sufficient at $K=4$ but not at $K=8$. Max-unlock similarly increases from $1.48\times$ to $2.00\times$; FIFO and LRU vary less with $K$ but remain substantially above the retained reference.

\subsection{Learning-Based Compilation}
\label{sec:nn}

Because optimal compilation is NP-hard, the effectiveness of learning-based routing depends critically on the quality of its supervision. Recent learning-based routers, such as the reinforcement-learning-based AlphaRouter~\cite{tang2024alpharouter}, have advanced this direction by demonstrating the significant potential of policy-driven compilation. Yet, even state-of-the-art learning-based methods remain fundamentally bounded by the quality and scalability of their training labels. In this work, we use learning-based routing not to introduce a new algorithm, but rather as a controlled probe to isolate and quantify the impact of supervision quality.
A useful supervision source should provide calibrated labels, transfer to real workloads, and remain informative as system size grows. We evaluate QROB under this criterion by training the \emph{identical} EnhancedGATPolicy architecture on different supervision sources, isolating supervision quality and workload conditioning from model capacity.

\begin{figure*}[htbp]
    \centering
    \includegraphics[width=\textwidth]{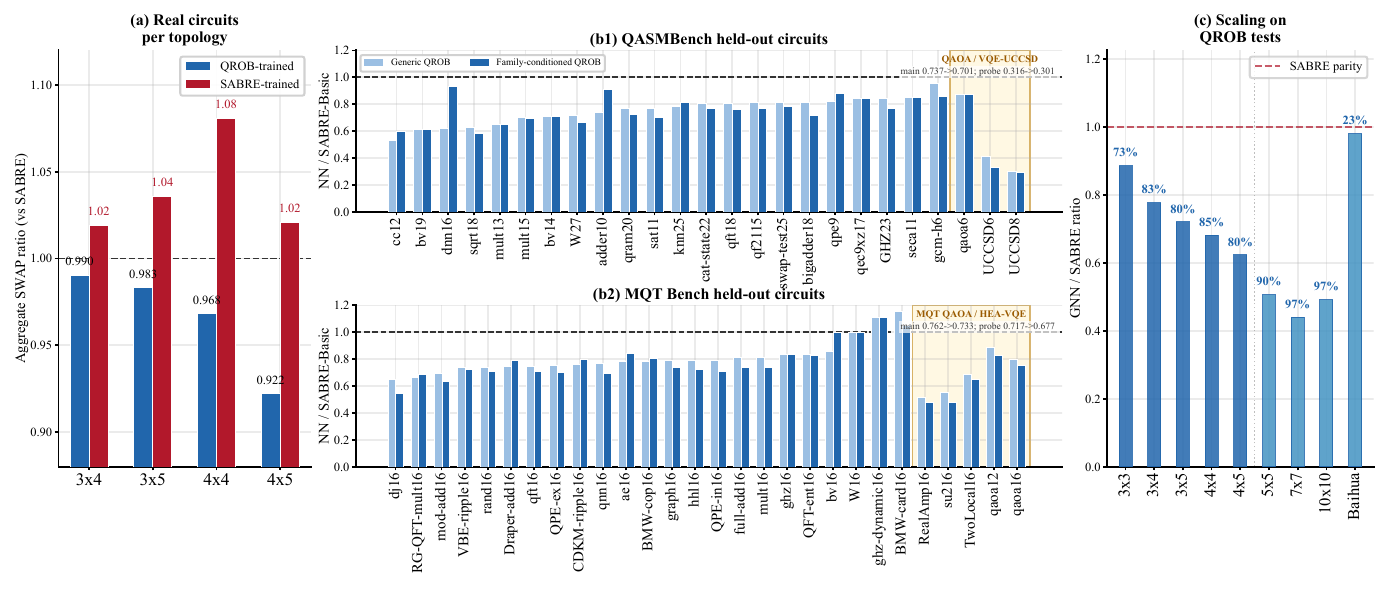}
    \caption{QROB references as supervision (IR mode) for learning-based routing.
    (a)~With the same policy architecture, QROB-supervised policies outperform SABRE-supervised policies on real-circuit evaluation settings.
    (b1)--(b2)~Simulation on real application circuits from QASMBench and MQT Bench: generic QROB supervision already achieves lower SWAP ratios than SABRE-Basic, and family-conditioned QROB supervision using the Extended Variational Strategy further reduces the ratios, including on the highlighted QAOA/VQE-style probes.
    (c)~The QROB-supervised policy scales across grid and Baihua topologies; bar heights report aggregate SWAP ratio against SABRE-Basic and percentages report per-circuit beat rate.}
    \label{fig:nn_combined}
\end{figure*}

In a controlled calibration with the model architecture fixed and only the supervision source changed, switching from SABRE heuristic labels to QROB references raises the SABRE beat rate from 57.8\% to 82.2\% on 750 test circuits and lowers the reference-cost ratio from 1.93 to 1.29 (reproducible across 5 seeds, std $<0.5\%$).

Beyond calibration, the learned policy must also transfer to real workloads. Fig.~\ref{fig:nn_combined}(a) shows that the QROB-trained model outperforms SABRE across all four real-circuit topologies (aggregate SWAP ratio 0.922--0.990), whereas the SABRE-trained model underperforms SABRE on every topology (1.02--1.08). On 125 real quantum circuits from 10 algorithm families, the QROB-trained model beats SABRE on 84.8\% of circuits with 100\% completion rate. In contrast, an identical model trained on adversarial benchmark labels reaches 94.7\% validation accuracy yet fails to complete 75\% of real circuit compilations, showing that label correctness alone is insufficient when the benchmark structure diverges from real workloads.

We next evaluate QROB-supervised routing on real application circuits rather than only on QROB-generated tests. Fig.~\ref{fig:nn_combined}(b1)--(b2) reports simulated routing results on QASMBench and MQT Bench circuits against SABRE-Basic. On the main application suites, generic QROB supervision already reaches aggregate SWAP ratios of 0.737 on QASMBench and 0.762 on MQT Bench, showing that QROB-generated supervision transfers to real benchmark circuits and improves over the standard SABRE-Basic baseline.

We then use the same real-circuit evaluation to test whether QROB can generate supervision closer to a target workload family in routing-relevant structure. The highlighted probes include QASMBench QAOA and UCCSD-style VQE circuits, as well as MQT QAOA and hardware-efficient VQE circuits such as RealAmplitudes, SU2, and TwoLocal. Using the modular construction mechanism of \S\ref{subsec:ideal}, we condition QROB's ideal construction on an Extended Variational Strategy---a bounded-degree interaction graph decomposed into executable matching layers---while leaving the reverse-disruption and repair-SWAP labeling unchanged. Family-conditioned QROB improves the main-suite ratios to $0.701$ and $0.733$, and improves the QAOA/VQE-style probe ratios from $0.316$ and $0.717$ to $0.301$ and $0.677$. This shows that QROB instances can be shaped toward real workload routing footprints without using benchmark circuits as routing-label sources.

Finally, Fig.~\ref{fig:nn_combined}(c) shows that the benefit of QROB supervision persists across scale on QROB-generated tests. The generic QROB-supervised policy remains below SABRE-Basic in aggregate SWAP count across grid sizes from $3{\times}3$ to $10{\times}10$, with the strongest gains on larger grids, and reaches near-parity on the 155-qubit Baihua topology without retraining. We report both aggregate SWAP ratio and per-circuit beat rate because a small number of large circuits can dominate total routing cost.

\subsection{Practical Validity}
\label{sec:validation}

The value of calibrated references depends on whether they reflect practical compilation behavior rather than synthetic-construction artifacts. We therefore examine QROB's practical validity from three complementary perspectives: whether its generated instances resemble real circuit-library workloads, whether QROB-supervised policies transfer to practical circuits, and whether the optimized cost metric is predictive of real hardware execution quality.

\begin{figure}[htbp]
      \centering
      \includegraphics[width=\columnwidth]{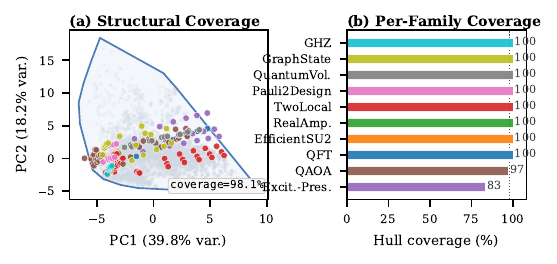}
      \caption{Structural representativeness of QROB-generated instances. (a)~PCA projection of 18 routing-relevant structural features. The convex hull of QROB instances covers 98.1\% of real-circuit points from 10 standard families in Qiskit's circuit library. (b)~Per-family containment in the same 2D PCA space; eight families are fully covered, while the only uncovered points appear in ExcitationPreserving and QAOA circuits, identifying boundary regimes for generic generation.}
      \label{fig:validation}
\end{figure}

We first examine workload-level validity. To assess whether QROB instances are structurally representative of real circuit-library workloads, we extract 18 structural features from 6,798 QROB circuits and 216 real quantum circuits from 10 standard families in Qiskit's circuit library, covering common ansatz, optimization, transform, randomized-benchmark, and graph/state-preparation structures: EfficientSU2, RealAmplitudes, TwoLocal, PauliTwoDesign, QAOA, ExcitationPreserving, QFT, QuantumVolume, GraphState, and GHZ. The 18 features span four routing-relevant groups---interaction-graph, circuit-structure, locality, and layer-pattern---that together characterize the structures determining mapping and routing difficulty. Fig.~\ref{fig:validation}(a) shows that the QROB convex hull in this 2D PCA space covers \textbf{98.1\%} of real-circuit points. Fig.~\ref{fig:validation}(b) reports per-family containment: eight of the ten families are fully covered, while the remaining uncovered points appear only in ExcitationPreserving circuits (83\% covered) and QAOA circuits (97\% covered). The few uncovered points are boundary cases rather than a broad missing workload class: one small QAOA instance whose interactions are already nearest-neighbor on the reference grid, and three largest-scale ExcitationPreserving instances in our set with dense all-to-all interaction graphs, long interaction distances, and strong layer imbalance.
These results indicate that QROB-generated instances closely span the routing-relevant feature regimes occupied by common real circuit-library workloads, while also identifying the few boundary regimes where generic generation is less complete.

The learning-transfer results of \S\ref{sec:nn} provide an operational counterpart to this structural check: QROB-supervised policies improve routing on held-out QASMBench and MQT Bench workloads, so structural coverage translates into downstream benefit. When a generic generator underrepresents a family, initializing $C_{\mathrm{ideal}}$ from that family's interaction footprint (\S\ref{sec:nn}) recovers it.

\begin{figure}[htbp]
    \centering
    \begin{minipage}{1.0\linewidth}
		\centerline{\includegraphics[width=\textwidth]{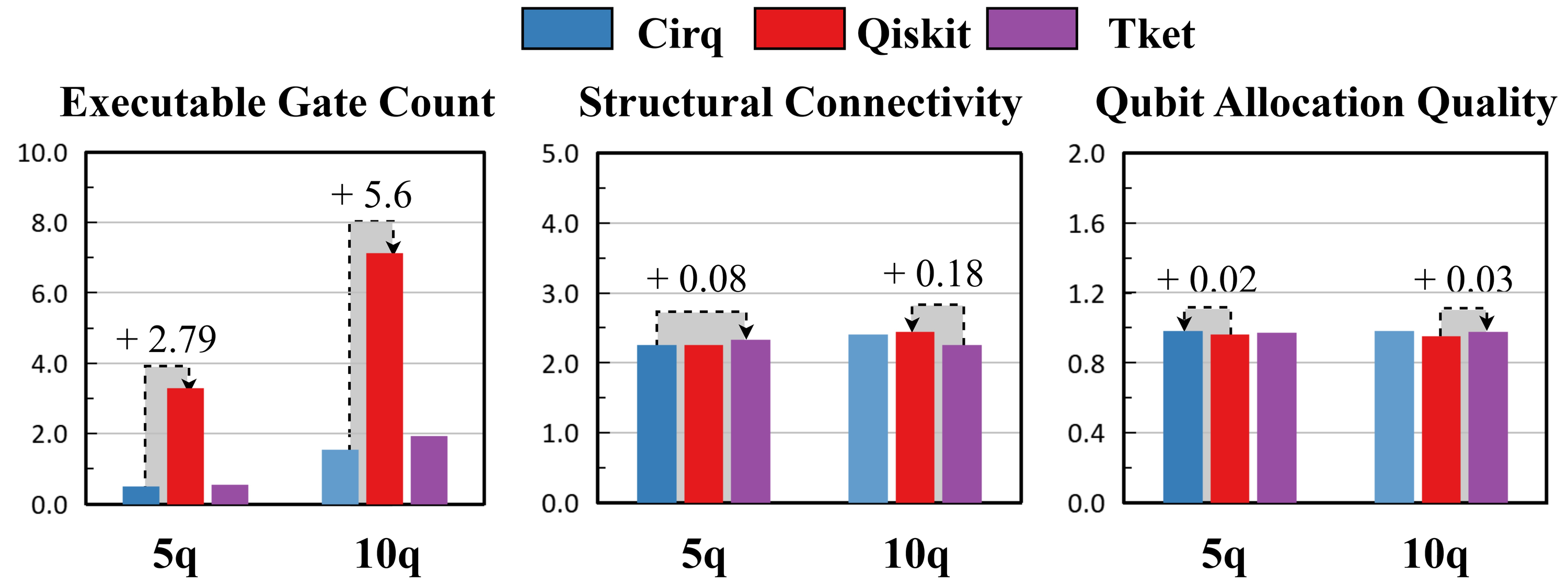}}
	\end{minipage}
	\begin{minipage}{1.0\linewidth}
		\centerline{\includegraphics[width=\textwidth]{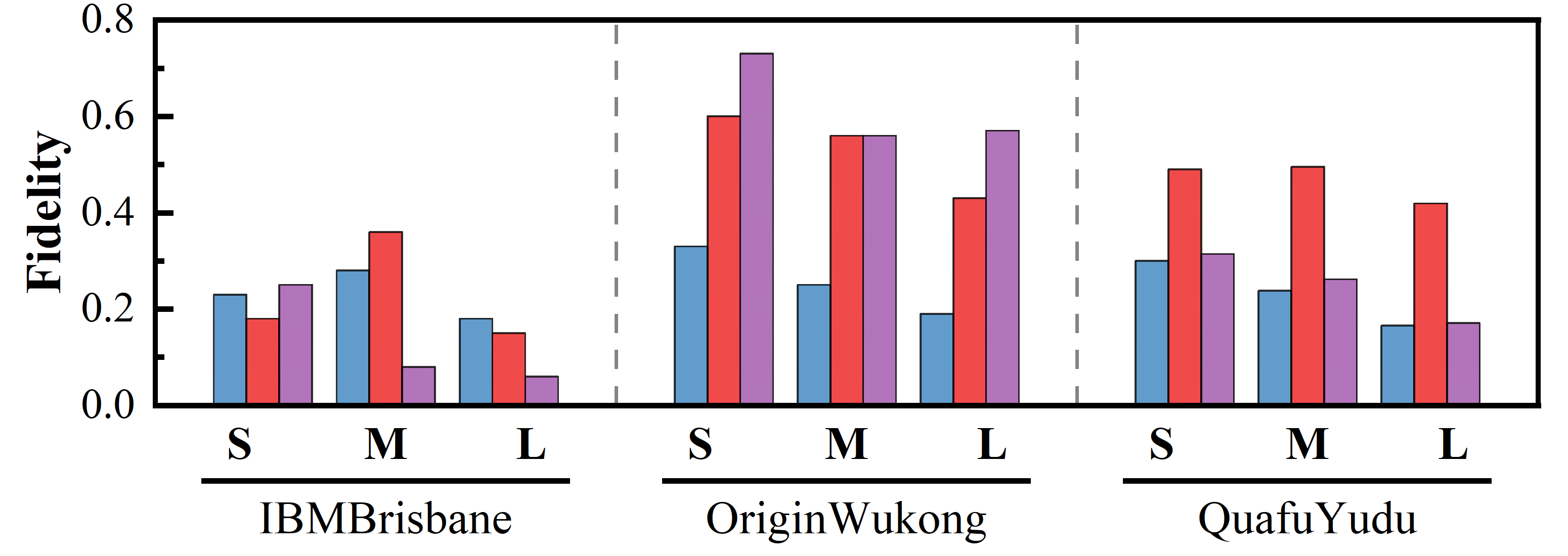}}
	\end{minipage}
    \caption{Multi-platform fidelity validation. Top: initial mapping analysis across three compilers. Bottom: execution fidelity on IBM Brisbane, Origin Wukong, and Quafu Yudu processors confirms consistent cost-fidelity correlation across platforms.}
    \label{fig:multiplatform}
\end{figure}

We next examine physical validity. On the Quafu Baihua 156-qubit processor, controlled-Z (CZ) gate count explains \textbf{85.4\%} of cross-entropy benchmarking (XEB) fidelity variance~\cite{cross2019validating} ($R^2{=}0.854$), confirming that the cost metric optimized by QROB is strongly predictive of actual hardware execution quality. This relationship is consistent with device-level gate physics: with per-gate fidelity $F_{CZ}{=}0.965$, each additional SWAP reduces fidelity by approximately $F_{SWAP}{\approx}0.898$. Fig.~\ref{fig:multiplatform} further shows the same cost--fidelity correlation across IBM Brisbane, Origin Wukong, and Quafu Yudu, indicating that the relevance of the metric is not specific to a single processor. These results support the use of QROB references not only as abstract optimality anchors, but as references for a quantity that materially affects execution quality on real hardware.

\subsection{Fidelity Gap under Full Industrial Optimization}\label{sec:o3gap}

\begin{figure}[htbp]
    \centering
    \includegraphics[width=\columnwidth]{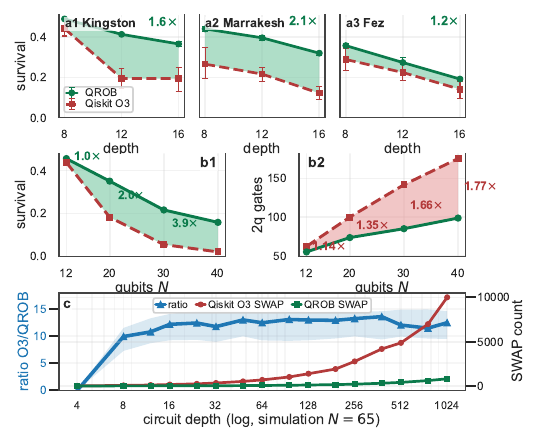}
    \caption{Real-hardware fidelity gap under full \texttt{Qiskit} O3 versus the QROB reference (IR mode), on three 156-qubit IBM Heron-r2 processors. \textbf{(a1--a3)}~per-device mirror-circuit survival vs.\ depth at $N{=}20$ on \textit{ibm\_kingston}, \textit{ibm\_marrakesh}, and \textit{ibm\_fez}; \textbf{(b1,b2)}~its growth with system size in fidelity and two-qubit-gate count (\textit{ibm\_marrakesh}); \textbf{(c)}~the O3/QROB routing-SWAP ratio across circuit depth in simulation ($N{=}65$; left axis ratio, right axis SWAP count). In-panel numbers are median QROB/O3 ratios.}
    \label{fig:hw_fidelity}
\end{figure}

\S\ref{sec:validation} established that the optimized routing cost correlates with hardware fidelity; we now measure the fidelity gap directly, comparing QROB references head-to-head against a full industrial pipeline on three 156-qubit IBM Heron-r2 processors (\textit{ibm\_kingston}, \textit{ibm\_marrakesh}, \textit{ibm\_fez}). Each instance is compiled through two controlled configurations that differ only in the source of the joint layout/routing solution: the O3 configuration runs a full noise-aware \texttt{Qiskit} O3 pass with free choice of layout and routing over all 156 qubits, while the QROB configuration fixes the QROB-supplied layout and routing and then applies the same backend-aware downstream O3 passes with re-layout and re-routing disabled. Both configurations share the same calibration snapshot, basis gates, scheduling, and shot budget, and fidelity is measured end-to-end as Pauli-randomized mirror-circuit survival~\cite{proctor2022mirror} over the same 20 logical qubits. As shown in Fig.~\ref{fig:hw_fidelity}(a1--a3), the QROB reference wins 57 of 65 instance pairs, with a combined median survival ratio of \textbf{1.65$\times$} (paired sign-test $p \approx 3\times10^{-10}$), reproducible across three generation seeds and reruns.

To locate the source, we compare the two realizations' resources. Both configurations execute on CZ edges of statistically indistinguishable fidelity (mean difference $\leq 2\times10^{-4}$)---O3's noise-aware layout also finds high-quality edges---yet the O3 realization uses more two-qubit gates. The survival gap therefore traces to this gate-count overhead, the joint-mapping optimality gap, and both grow with system size: on \textit{ibm\_marrakesh} the two-qubit-gate ratio rises from 1.14$\times$ at $N{=}12$ to 1.77$\times$ at $N{=}40$, and the median survival ratio correspondingly grows to 3.9$\times$ at $N{=}30$ (Fig.~\ref{fig:hw_fidelity}(b1,b2)). A survival ratio is bounded and therefore lives on a different scale from the unbounded SWAP-count ratios of \S\ref{sec:gap}; the overhead is paid in fidelity and compounds with scale. The experiment is an existence demonstration on QROB-generated instances---constructing instances whose near-optimal joint mapping the industrial pipeline does not find is precisely the benchmark's purpose.

The same headroom appears in transpilation alone and holds across circuit depth (Fig.~\ref{fig:hw_fidelity}(c)): enabling full downstream O3 reduces but does not close it, and the hardware results above show that what remains is paid directly in fidelity.

\subsection{Summary}

Calibrated references reveal a compilation overhead that is large ($7.2\times$--$24\times$ on coupling maps, up to $6.3$--$7.0\times$ on surface-code lattices), grows with system scale, and persists across compilation layers. 
Separately, the QROB-Mem extension shows that a fixed absolute scheduling horizon does not transfer across compute-region capacities, whereas the evaluated $2K$-gate next-use horizon remains close to the retained reference. As a supervision source, QROB references enable learned compilers to outperform SABRE on 84.8\% of real circuits, transfer to held-out QASMBench and MQT Bench workloads, and support family-conditioned supervision for QAOA/VQE-style probes without using benchmark circuits as routing-label sources. 
Validation confirms that QROB measures the right quantity: generated instances cover 98.1\% of real-circuit points in a routing-relevant feature space, the optimized metric explains 85\% of hardware fidelity variance, compiler rankings transfer reliably to practical circuits, and the median QROB/O3 mirror-circuit survival ratio is $1.65$ across three IBM Heron-r2 processors.

\section{Related Work}
\label{sec:related}
Exact formulations of qubit mapping and routing~\cite{tan2020optimal, lin2023scalable, zhang2021time} certify optimality on small instances, but do not scale beyond a few dozen qubits. Practical toolchains therefore rely on heuristics such as SABRE and LightSABRE~\cite{li2019tackling, zou2024lightsabre}, \tket~\cite{sivarajah2020tket}, and MQT QMAP~\cite{zulehner2018efficient}. These handle hundreds of qubits, but report only the cost they reach, with no reference to measure it against. Learning-based routers~\cite{tang2024alpharouter, pozzi2022using, ostaszewski2021reinforcement} train policies on labels or rewards produced by the same heuristics, so their quality is bounded by the quality of those labels. Noise- and ISA-aware routers~\cite{murali2019noise, yang2026reqisc, yang2026canopus} change what is optimized, folding in gate fidelities or richer gate sets, but face the same measurement question. QROB constructs instances together with feasible realizations, so all of these routers can be measured against a known solution, and the learning-based ones can draw labels from a source that is not itself a heuristic.

Benchmarks show a similar gap. Application suites~\cite{quetschlich2023mqt, tomesh2022supermarq, li2023qasmbench, nation2025benchpress}, their fault-tolerant counterpart FTCircuitBench~\cite{ftcircuitbench}, and analysis frameworks such as QuCT~\cite{tan2023quct} supply realistic circuits but no reference cost, so compilers can only be ranked against one another on them. Constructive benchmarks address this by building a known solution into each instance; \queko~\cite{tan2021optimality}, QUBIKOS~\cite{ping2025qubikos}, and \qknob~\cite{kim2024qknob} do so for SWAP routing on coupling maps. QROB generates its instances by scrambling, which spans a wider range of circuit structures and can be steered toward a target workload family, and covers both coupling-map routing and lattice-surgery scheduling with one construction. The quantum hardware roofline of Kalloor et al.~\cite{kalloor2024quantum} is a different kind of reference: it sets per-workload ceilings, but derives them from compiled outputs, so it cannot separate headroom left by the compiler from limits set by the hardware.

In the fault-tolerant setting, lattice-surgery compilers such as WISQ/DASCOT~\cite{molavi2025wisq}, EDPC~\cite{beverland2022edpc}, LaSsynth~\cite{tan2024lassynth}, AutoBraid~\cite{hua2021autobraid}, and others~\cite{chamberland2022universal, watkins2024high} optimize patch placement and channel scheduling under surface-code constraints, and have so far been compared only with one another. We evaluate WISQ/DASCOT and EDPC against QROB references. Distillation-aware layout and routing~\cite{sharma2026spacetime} addresses magic-state supply, a different bottleneck from the routing studied here.

Compute/memory-separated architectures differ in how they organize storage and computation: qSIEVE introduces a quantum memory hierarchy, HetEC combines heterogeneous error-correcting codes across compute and memory regions, and LSQCA adopts a load/store architecture with computational registers and scan-access memory~\cite{viszlai2026qsieve, stein2025hetec, kobori2025lsqca}. Their codes and transfer mechanisms differ, but all must decide which logical qubits reside in a limited compute region. QROB-Mem keeps only this constraint, at the resident-set level.

\section{Conclusion}
\label{sec:conclusion} 

This work began from a simple observation: quantum compilation has long lacked calibrated reference points for the implementation-cost gap induced by architecture constraints. We addressed this gap with QROB, a reverse-construction methodology that constructs compilation instances backward from retained feasible realizations rather than searching forward for optima. In its core settings, QROB provides a common measurement framework for physical-qubit SWAP routing and fault-tolerant lattice-surgery scheduling, both of which reconcile interaction requirements with geometric routing constraints. QROB-Mem broadens this framework to capacity-constrained memory-access scheduling, illustrating its applicability beyond the two core geometric-routing settings.

Deploying this framework reveals that the compilation overhead is substantially larger, more persistent, and more consequential than previously observable: the overhead ratio grows with system size, persists across both near-term and fault-tolerant layers, and translates into a measured fidelity forfeiture on real hardware under full industrial optimization. 
For the evaluated next-use policy, the QROB-Mem extension further shows that a $2K$-gate lookahead maintains near-reference quality, whereas a fixed absolute horizon does not transfer across compute-region capacities.

Taken together, these results suggest that quantum compilation should be evaluated not only by relative comparisons among heuristics, but by how far practical compilers remain from calibrated reference costs. QROB provides a common measurement substrate for that purpose, supporting compiler evaluation, learning-based routing, and hardware-relevant diagnosis within the same reverse-construction framework. We will release the methodology and benchmark suite upon publication to support more reproducible and quantitatively grounded progress in quantum compilation.

More broadly, the reverse-construction perspective may provide a useful foundation for studying additional architecture-constrained compilation settings beyond those considered here.

\section*{Acknowledgment}

We thank the anonymous reviewers for their constructive feedback. 
This work is supported by the Quantum Science and Technology--National Science and Technology Major Project (Grant Nos. 2024ZD0300500, and 2021ZD0301800), 
National Natural Science Foundation of China (Grant Nos. U25A6009, 92265207, 12247168, 12404578, 92365301, T2121001, and 92565391), 
MOST of China (2025YFE0217600), 
Beijing National Laboratory for Condensed Matter Physics (2024BNLCMPKF022), 
Young Elite Scientists Sponsorship Program of the Beijing High Innovation Plan (Grant No. 20250945), 
Beijing Natural Science Foundation (Grant No. 1262048). 

\appendix
\section{Artifact Appendix}

\subsection{Abstract}

This artifact packages the QROB reverse-construction framework and reproduces
the paper's results as six self-contained sub-artifacts (containerized with
Docker), one per results section:
\textbf{coupling\_gap} (\S\ref{sec:gap}: Table~\ref{tab:compiler_gap} and Fig.~\ref{fig:overall}, from
one-minute aggregation up to an end-to-end
seed$\rightarrow$circuits$\rightarrow$routing$\rightarrow$numbers workflow);
\textbf{olsq2} (\S\ref{subsec:optimality}: solver-based validation on freshly generated
instances); \textbf{ftqc} (\S\ref{sec:ftqcgap}: re-executing WISQ/DASCOT and
EDPC/TeleportRouter); \textbf{nn} (\S\ref{sec:nn}--\ref{sec:validation}: learning-based compilation
from pretrained checkpoints with the full QROB test sets bundled, plus the
PCA representativeness analysis); \textbf{fidelity\_multiplatform} (\S\ref{sec:validation})
and \textbf{fidelity\_gap} (\S\ref{sec:o3gap}: hardware validation).
Every sub-artifact ships a \texttt{README}, the exact expected output to
check against, and a one-command entry point. Deterministic checks are
expected to match the archived outputs exactly under the provided
environments; stochastic and hardware-derived results carry explicitly
documented tolerances. No live quantum-hardware access is required:
hardware-backed results are reproduced by reanalyzing the archived
per-circuit and raw-device measurement records, and the collection scripts
are included with all credentials and access tokens removed.

\subsection{Artifact check-list (meta-information)}

{\small
\begin{itemize}[leftmargin=*, itemsep=0pt]
  \item \textbf{Algorithm:} QROB reverse construction (influence-guided
    scrambling with retained reference solutions).
  \item \textbf{Program:} QROB benchmark suites (shipped); QASMBench subset
    (shipped, BSD) and MQT Bench (generated on the fly); WISQ/DASCOT
    (Apache-2.0, included); EDPC/TeleportRouter (Microsoft non-commercial
    license, fetched at build time from its upstream repository at a pinned
    commit).
  \item \textbf{Compilation:} pinned stacks --- \texttt{qiskit} 1.2.4 (SABRE),
    \texttt{cirq} 1.5.0, \texttt{pytket} 2.7.0, \texttt{mqt.qmap} 2.6.0,
    \texttt{olsq} 0.0.4.1 + z3; Julia 1.9 (EDPC only).
  \item \textbf{Model:} seven pretrained EnhancedGATPolicy checkpoints (GATv2,
    47K parameters, $\sim$200\,KB each).
  \item \textbf{Data set:} 15-config coupling-map benchmark (280 circuits each,
    with per-circuit reference labels); lattice-surgery instances (64--256
    logical qubits); QROB learning test sets ($\sim$1.4\,GB, small grids
    through the 155-qubit Baihua topology); frozen aggregates and feature
    tables; recorded hardware measurements (IBM, Origin, Quafu; tokens
    removed).
  \item \textbf{Run-time environment:} Docker (containerized environments per
    sub-artifact; \texttt{coupling\_gap} ships an aggregation image and a
    pinned-router image; Linux/x86-64).
  \item \textbf{Hardware:} any x86-64 machine, $\ge$16\,GB RAM; no GPU or
    quantum hardware required for any evaluated claim.
  \item \textbf{Metrics:} SWAP/depth/time overhead ratios; lattice-surgery
    makespan and space--time volume; routing win-rate and SWAP ratio; PCA
    convex-hull coverage; circuit fidelity and mirror-circuit survival.
  \item \textbf{Output:} console tables matching each sub-artifact's
    documented expected results; regenerated figures where claimed.
  \item \textbf{Experiments:} one documented entry point per sub-artifact
    (\texttt{run\_*.sh}/\texttt{reproduce.sh}, or a stdlib Python script for
    \S\ref{sec:o3gap}); quick and, where applicable, full modes.
  \item \textbf{Code licenses:} author code MIT; WISQ/DASCOT Apache-2.0
    (included); EDPC engine Microsoft non-commercial (fetched at build,
    not redistributed).
  \item \textbf{Data licenses:} author-generated data and documentation
    CC~BY~4.0; QASMBench subset BSD (license text included).
  \item \textbf{Workflow automation framework used:} Docker + Bash entry
    scripts.
  \item \textbf{Disk space:} $\sim$10\,GB (images + data).
  \item \textbf{Time to prepare:} $\sim$30\,min (Docker builds).
  \item \textbf{Time to complete:} $\sim$1--2\,h for all quick paths; optional
    full re-runs (hours--days) are documented per script.
  \item \textbf{Publicly available:} yes (Zenodo).
  \item \textbf{Archived:} DOI: \texttt{10.5281/zenodo.21539550}.
\end{itemize}
}

\subsection{Description}

\subsubsection{How to access}
Archived at \url{https://doi.org/10.5281/zenodo.21539550} (also provided to
reviewers through the submission site). Six self-contained sub-directories:
\texttt{coupling\_gap/}, \texttt{olsq2/}, \texttt{ftqc/}, \texttt{nn/},
\texttt{fidelity\_multiplatform/}, \texttt{fidelity\_gap/}, each with its own
\texttt{README.md}, documented expected results, \texttt{Dockerfile}, and run
scripts.

\subsubsection{Hardware dependencies}
Commodity x86-64; $\ge$16\,GB RAM recommended. No GPU and no quantum-hardware
access are required for any evaluated claim; the original hardware experiments
ran on IBM Heron-r2 (156q), IBM Brisbane, Origin Wukong, Quafu Yudu, and Quafu
Baihua, whose recorded measurements ship with the artifact.

\subsubsection{Software dependencies}
Docker. Internet access is needed only while building the images (pinned
package downloads; the \texttt{ftqc} build additionally fetches a pinned
upstream revision of TeleportRouter) --- the experiments themselves run
offline. (Without Docker: Python 3.8--3.12 per-sub-artifact
\texttt{requirements.txt}; Julia 1.9 for EDPC only. The \S\ref{sec:o3gap} reproduction
path needs the Python standard library only.)

\subsubsection{Data sets}
All inputs ship inside the archive, including the complete coupling-map
benchmark with retained per-circuit reference labels, the frozen per-config
aggregate behind Table~\ref{tab:compiler_gap}, the learning figure's full QROB test sets
($\sim$1.4\,GB, small grids through the 155-qubit Baihua topology), and the
recorded hardware measurements (Quafu raw device records included verbatim
with access tokens removed).

\subsection{Installation}
{\footnotesize
\begin{verbatim}
# per sub-artifact, e.g.:
cd coupling_gap
docker build -t qrob-coupling .
# quick path:
docker run --rm qrob-coupling
# deeper levels:
docker build -f Dockerfile.l2 -t qrob-l2 .
\end{verbatim}
}

\subsection{Experiment workflow}

Each sub-artifact is one \texttt{docker run} for the fast reproduction, with
deeper optional scripts documented in its \texttt{README}:
\texttt{coupling\_gap} layers \texttt{run\_all} (aggregation, $<$1\,min),
\texttt{run\_l2} (pinned re-routing), \texttt{run\_l1} (seed-42 regeneration)
and \texttt{run\_e2e}; \texttt{olsq2/reproduce.sh} solves fresh IR-mode
instances live; \texttt{ftqc/run\_all.sh} re-executes both lattice-surgery
compilers; \texttt{nn/reproduce.sh} re-evaluates the pretrained checkpoints,
verifies the recorded real-circuit transfer, and recomputes the PCA coverage;
the two fidelity sub-artifacts recompute every claimed number from the
recorded measurements.

\subsection{Evaluation and expected results}

\textbf{Key results} (each printed by the listed entry point and matched
against the sub-artifact's documented expected results):
\begin{itemize}[leftmargin=*, itemsep=0pt]
  \item \emph{Table~\ref{tab:compiler_gap}} (all four columns): SWAP 7.2/12.9/16.0/23.0$\times$,
    depth, time, and scale reproduced to $\pm$0.1; the
    3$\times\!\rightarrow\!$24$\times$ scale statement; and
    Fig.~\ref{fig:overall}'s twelve per-config ratio annotations
    ($\times$1.48\,\ldots\,$\times$24.12) reproduced exactly.
  \item \emph{L2 re-routing}: per-circuit SWAP counts of the six deterministic
    routers match the shipped logs \textbf{exactly} (100\%) under the provided
    Docker environment (verified on Intel and AMD hosts);
    \texttt{stochastic\_swap} means agree within 2\%.
  \item \emph{L1 regeneration}: seed-42 regeneration is content-identical to
    the shipped dataset (ankaa9Q: 560/560 files, 280/280 labels).
  \item \emph{OLSQ2}: fresh IR-mode instances give QROB reference $=$ depth
    exactly; ordering OLSQ2 $\le$ QROB $\le$ SABRE $\le$ Basic/Stochastic;
    OLSQ2 means within $\sim$0.3 mean SWAPs of the published data at
    fully-solved depths; the high-depth timeout wall reproduces the
    scalability claim (per-depth solved counts documented in
    \texttt{EXPECTED.md}).
  \item \emph{FTQC}: WISQ 2.1$\rightarrow$6.3$\times$ and EDPC
    3.4$\rightarrow$7.0$\times$ time overhead (L=64$\rightarrow$256); volume
    6.4--6.5$\times$ at L=256; compact layout 9.2$\rightarrow$43.8$\times$;
    density endpoints 3.9$\times$/4.4$\times$; schedule verification prints
    \texttt{VERDICT: PASS} (reference 4 steps vs.\ WISQ 10). Fresh re-runs
    reproduce the per-L means within a few percent (8 instances per point;
    simulated annealing is stochastic).
  \item \emph{Learning}: from the pretrained checkpoints, QASMBench/MQT
    aggregate SWAP ratios 0.737/0.762 (generic) and 0.701/0.733
    (family-conditioned), probes 0.316$\rightarrow$0.301 and
    0.717$\rightarrow$0.677, re-evaluated within rollout noise ($<$1\%); the
    84.8\%-of-125 real-circuit transfer recomputed from shipped per-circuit
    records and re-runnable end-to-end (expected 84.8\%, stochastic rollouts;
    our verification re-run: 82.4\%);
    the 57.8\%$\rightarrow$82.2\% calibration ships as the recorded evaluation
    and re-runs from the bundled test set (\texttt{nn/code/scaling/}).
  \item \emph{Representativeness}: PCA convex-hull coverage 98.1\%, 8/10
    families fully covered (recomputed from the shipped feature CSVs,
    deterministic; a from-zero circuit-generation$+$feature-extraction
    pipeline is also included).
  \item \emph{Multi-platform fidelity}: 18/27 platform$\times$size$\times$compiler
    cells reproduce exactly from the per-circuit records; the 9 Quafu Yudu
    cells derive from raw device records within $|\Delta|\le 0.011$; the
    Baihua CZ$\rightarrow$XEB fit ($R^2{=}0.854$, slope $-0.0358$) refits
    exactly from the recorded 93 points.
  \item \emph{Fidelity gap (\S\ref{sec:o3gap})}: combined survival median 1.65$\times$ ---
    clean 57/65 (equal-cardinality pairs where both arms measured all 20 logical
    qubits; 7 excluded where the O3 layout left a qubit idle, \texttt{nmeas}
    17--19), median 1.646$\times$, sign-test $p{=}3.16\times10^{-10}$; raw
    62/72, 1.638$\times$ --- the script prints both; marrakesh scaling 1.14$\rightarrow$1.77$\times$
    (2q gates) and 3.9$\times$@$N{=}30$ (survival); CZ-edge difference
    $\le 2\times10^{-4}$; simulated O3/QROB ratio $\sim$12--13$\times$;
    multi-seed pooled 1.49$\times$, 69/72 --- all recomputed from the shipped
    measurements by a stdlib-only script.
\end{itemize}
Known deviations are documented in the respective \texttt{README} files
(notably: the 156-qubit headline point reproduces from the paper's frozen
measurement record, whose reference cell averages over the 208 Cirq-routed
circuits; a ratio over all 280 circuits computes $\approx$16$\times$ --- every
quantity is recomputable from the shipped per-circuit files; see
\texttt{coupling\_gap/README.md}~\S7).

\subsection{Experiment customization}

The QROB generator is fully parameterized (chip/topology, seed, size and
density presets); routers accept per-algorithm configs; the OLSQ2 driver
exposes parallelism, timeout, and depth range; the FTQC sweep and NN
evaluations expose instance-size and model-selection knobs (see each
\texttt{README}).





\bibliographystyle{IEEEtran}
\bibliography{reference}

\end{document}